\pdfoutput=1
\documentclass{article}

\PassOptionsToPackage{table}{xcolor}
\usepackage{report}

\usepackage{latexsym}
\usepackage[T1]{fontenc}
\usepackage[utf8]{inputenc}

\usepackage{xspace}
\usepackage{graphicx}
\usepackage{dblfloatfix}
\usepackage{subcaption} 
\usepackage{booktabs}
\usepackage{listings}
\usepackage{placeins}

\usepackage{amsmath}
\usepackage{amssymb}
\usepackage{mathtools}

\usepackage[capitalize,noabbrev]{cleveref}

\usepackage[most]{tcolorbox}
\usetikzlibrary{positioning,arrows.meta}
\usepackage{forest}

\newcommand{\system}{\textsc{CodeTracer}\xspace}
\newcommand{\bench}{\textsc{CodeTraceBench}\xspace}
\newcommand{\miniSystem}{\textsc{Mini-CodeTracer}\xspace}
\newcommand{\figMainWide}{0.75\linewidth}
\newcommand{\figMainMedium}{0.70\linewidth}
\newcommand{\figGridIterW}{0.26\textwidth}
\newcommand{\figGridRangeW}{0.27\textwidth}
\newcommand{\figCardTwoW}{0.54\linewidth}

\newcommand{\parabf}[1]{\noindent\textbf{#1}\xspace}

\tcbset{
  takeaway/.style={
    width=\linewidth,
    top=6pt,
    bottom=4pt,
    left=4pt,
    right=4pt,
    colback=white,
    colframe=black,
    colbacktitle=black,
    coltitle=white,
    fonttitle=\bfseries\small,
    enhanced,
    boxrule=0.6pt,
    toptitle=2pt,
    bottomtitle=2pt,
  }
}

\newtcolorbox{takeawaybox}[1]{takeaway,title={#1}}

\tcbset{
  promptbox/.style={
    enhanced,
    breakable,
    colback=red!4!yellow!4!white,
    colframe=red!25!yellow!20!white,
    colbacktitle=red!18!yellow!12!white,
    coltitle=black,
    fonttitle=\bfseries\small,
    boxrule=0.4pt,
    arc=2mm,
    left=3mm, right=3mm, top=2mm, bottom=2mm,
    attach boxed title to top left={yshift=-2mm, xshift=3mm},
    boxed title style={arc=1mm, boxrule=0.3pt},
  }
}

\title{CodeTracer: Towards Traceable Agent States}
\author{
\textbf{Han Li}$^{1}$\,\thanks{Equal contribution.}\quad
\textbf{Yifan Yao}$^{1}$\,\footnotemark[1]\quad
\textbf{Letian Zhu}$^{1}$\,\footnotemark[1]\quad
\textbf{Rili Feng}$^{1}$\,\footnotemark[1]\quad
\textbf{Hongyi Ye}$^{1}$\quad
\textbf{Jiaming Wang}$^{1}$\\
\textbf{Yancheng He}$^{6}$\,
\textbf{Pengyu Zou}$^{1}$\,
\textbf{Lehan Zhang}$^{5}$\,
\textbf{Xinping Lei}$^{1}$\,
\textbf{Haoyang Huang}$^{2}$\\
\textbf{Ken Deng}$^{2}$\,
\textbf{Zizheng Zhan}$^{2}$\,
\textbf{Ming Sun}$^{2}$\,
\textbf{Zhaoxiang Zhang}$^{3}$\,
\textbf{He Ye}$^{4}$\,
\textbf{Jiaheng Liu}$^{1}$\,\thanks{Corresponding author.}\\[2pt]
\vspace{3mm}
\normalsize
$^{1}$\,\textbf{Nanjing University}\quad
$^{2}$\,\textbf{Kuaishou Technology}\quad
$^{3}$\,\textbf{Institute of Automation, CAS}\quad \\
$^{4}$\,\textbf{University College London}\quad
$^{5}$\,\textbf{Renmin University of China}\quad
$^{6}$\,\textbf{Alibaba Group} \\
\vspace{2mm}
\texttt{han.li.cs@smail.nju.edu.cn},
\texttt{liujiaheng@nju.edu.cn} \\
}
\begin{document}
\maketitle

\begin{abstract}
Code agents are advancing rapidly, but debugging them is becoming increasingly difficult. As frameworks orchestrate parallel tool calls and multi-stage workflows over complex tasks, making the agent's state transitions and error propagation hard to observe. In these runs, an early misstep can trap the agent in unproductive loops or even cascade into fundamental errors, forming hidden error chains that make it hard to tell \emph{when} the agent goes off track and \emph{why}. Existing agent tracing analyses either focus on simple interaction or rely on small-scale manual inspection, which limits their scalability and usefulness for real coding workflows. We present \system, a tracing architecture that parses heterogeneous run artifacts through evolving extractors, reconstructs the full state transition history as a hierarchical trace tree with persistent memory, and performs \textbf{failure onset localization} to pinpoint the failure origin and its downstream chain. To enable systematic evaluation, we construct \bench from a large collection of \emph{executed} trajectories generated by four widely used code agent frameworks on diverse code tasks (e.g., bug fixing, refactoring, and terminal interaction), with supervision at both the stage and step levels for failure localization. Experiments show that \system substantially outperforms direct prompting and lightweight baselines, and that replaying its diagnostic signals consistently recovers originally failed runs under matched budgets. Our code and data are publicly available.\footnote{\url{https://github.com/NJU-LINK/CodeTracer}}
\end{abstract}
\section{Introduction}

Large language models (LLMs) are increasingly used to power code agents that autonomously interact with software repositories and development environments~\citep{NEURIPS2024_5a7c9475,ICLR2025_a4b6ad6b,pmlr-v235-wang24h,xia2025demystifying}. These agents execute long sequences of heterogeneous actions—searching code, reading files, editing implementations, running builds, and interpreting test feedback—to solve complex engineering tasks such as repository-level bug fixing or system configuration~\citep{ICLR2024_edac78c3,merrill2026terminalbench}. As the capability of such agents improves, their executions grow longer and more complex, making understanding failures increasingly difficult. For example, when an agent run fails, it is often unclear where the trajectory first went wrong or which intermediate decisions caused the final failure.

Current evaluation practices provide only limited visibility into these processes. Most benchmarks summarize agent performance using end-to-end metrics such as pass rate or patch correctness, collapsing entire trajectories into a single success or failure label~\citep{ICLR2024_edac78c3,merrill2026terminalbench,xia2025demystifying}. 
Existing analyses of agent trajectories either rely on coarse outcome-level judgments or manual inspection of small numbers of runs, making them difficult to scale to realistic software engineering workloads with long execution traces~\citep{bouzenia2025trajectories,enconda2025}.

In this work, we introduce \system, a framework for converting heterogeneous agent run directories into structured hierarchical traces and automatically identifying the earliest failure critical stage in a trajectory. 
The system can also produce actionable debugging signals that can be fed back into the agent through reflective replay, enabling targeted recovery from earlier mistakes~\citep{shinn2023reflexion,madaan2023selfrefine,chen2024selfdebug}.
To support systematic evaluation, we construct \bench, a high-quality benchmark of code agent trajectories with step-level annotations. The benchmark aggregates trajectories generated by multiple widely used agent frameworks across diverse software engineering workloads, including repository level bug fixing and long horizon terminal interaction tasks~\citep{ICLR2024_edac78c3,NEURIPS2024_5a7c9475}. Each trajectory is annotated with structured step metadata and failure critical labels, enabling evaluation of both stage level localization and evidence retrieval.

Using this benchmark, we conduct a large-scale empirical study of code agent behavior across multiple model backbones and agent frameworks. Our analysis reveals several systematic patterns, including an evidence-to-action gap, where agents often retrieve relevant information but fail to translate it into correct state changing actions; significant variance in action efficiency across trajectories even for strong models; and stage-dependent error modes that concentrate failure-critical decisions in specific phases of the workflow.


Overall, the contributions of this work are as follows:
(1)~\system, a scalable framework for hierarchical trajectory tracing and failure onset localization in code agent executions.
(2)~\bench, a benchmark of thousands of annotated code agent trajectories enabling systematic evaluation of process level diagnosis.
(3)~A large-scale empirical analysis of failure patterns in modern code agents, revealing actionable insights into their reasoning and execution behavior.

\begin{figure*}[t]
    \centering
    \includegraphics[width=1.0\linewidth]{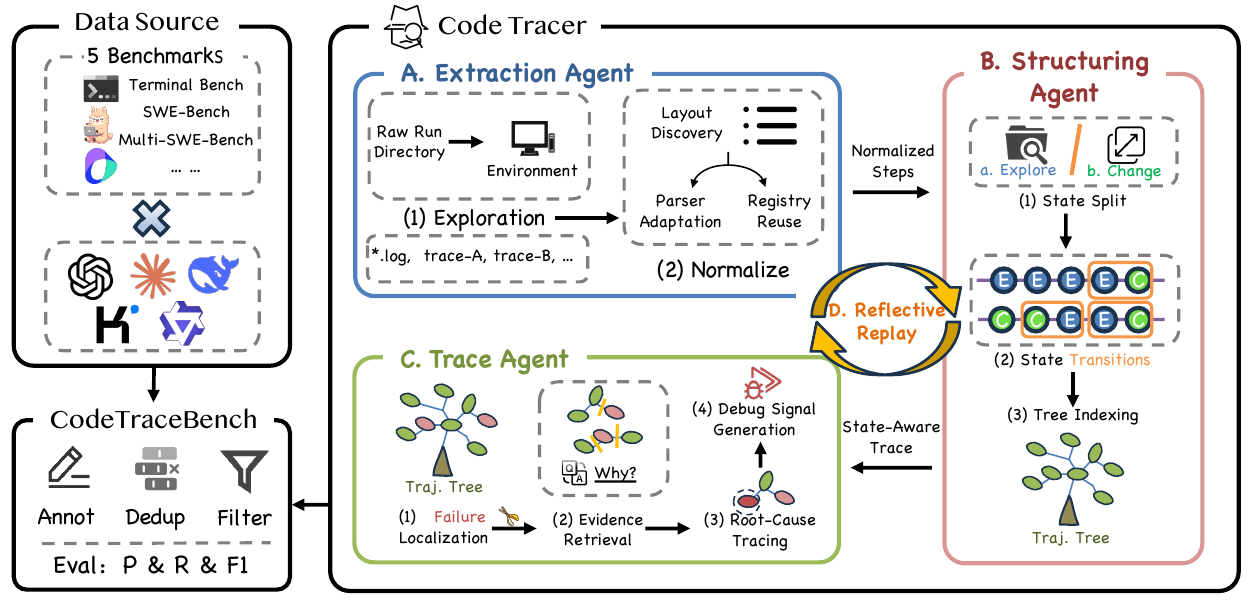}
    \caption{\textbf{Overview of the \system pipeline.} Raw trajectories are standardized into hierarchical traces, curated into \bench with step-level supervision, and diagnosed via failure onset localization with optional reflective replay.}
    \label{fig:overview}
\end{figure*}

\section{Trajectory Analysis}
\label{sec:study}

\subsection{Trajectory Collection}
We broadly collect execution trajectories from benchmarks that evaluate code generation and software engineering capabilities, covering both repository-level bug fixing (\textit{SWE-bench Verified}, \textit{SWE-bench Pro}, \textit{MultiSWE-bench}, \textit{SWE-PolyBench}) and long horizon terminal interaction (\textit{TerminalBench})~\citep{NEURIPS2024_5a7c9475,ICLR2024_edac78c3,ICLR2025_a4b6ad6b,merrill2026terminalbench}.
Each benchmark is executed under four widely used agent frameworks (\textit{SWE-Agent}, \textit{MiniSWE-Agent}, \textit{OpenHands}, \textit{Terminus~2})~\citep{NEURIPS2024_5a7c9475,ICLR2025_a4b6ad6b,merrill2026terminalbench} paired with five model backbones (\textit{Claude-sonnet-4}, \textit{GPT-5}, \textit{DeepSeek-V3.2}, \textit{Qwen3-Coder-480B}, \textit{Kimi-K2-Instruct})~\citep{deepseekai2025deepseekv32pushingfrontieropen,anthropic2025sonnet,kimiteam2026kimik2openagentic,qwen_qwen3_coder_next_tech_report,openai2025gpt5}, yielding a diverse corpus that spans different task regimes, orchestration strategies, and backbone capabilities.

\subsection{Filtering}
To obtain a consistent unit of progression across heterogeneous frameworks, we normalize each run into \emph{iterations} defined by executed commands and apply quality-based filtering in this normalized space.
Starting from 7{,}936 raw trajectories, we apply four successive filters:
(i)~removing runs that time out before completion (retaining 6{,}511);
(ii)~removing runs with incomplete or truncated generation traces (6{,}109);
(iii)~removing runs whose Docker environment is misconfigured or whose task files are corrupted, rendering the execution output unreliable (5{,}284);
(iv)~removing \emph{correct} trajectories with fewer than 10 normalized steps, which typically complete trivially and provide limited signal for failure analysis (3{,}326).
After filtering, we retain 3{,}326 trajectories spanning all benchmark--framework--backbone combinations.

\subsection{Annotation}
\label{sec:annotation}

All annotations are performed by the authors of this paper.
Each annotator is assigned a set of \emph{tasks} together with \emph{all} corresponding trajectories across the full backbone--agent grid (15 groups per task), and is provided with the task specification, the reference solution, and access to the execution environment for manual verification of ambiguous cases (e.g., environment or configuration issues).
This allocation allows each annotator to build deep familiarity with the task, facilitates horizontal comparison of how different models and agents approach the same problem, and enables efficient deduplication of similar failure modes.

For every trajectory, the annotator assigns a \emph{stage label} to each step, indicating the phase of the workflow (environment verification, dependency installation, inspection/debugging, patching, verification).
For \emph{successful} trajectories (i.e., those that pass all tests), the annotator further identifies \emph{redundant} and \emph{trial-and-error} steps—actions that do not contribute to the final solution or that are later reverted.
For \emph{failed} trajectories, we adopt a \emph{chain based backward tracing} protocol: starting from the failing test output, the annotator identifies the immediately preceding step whose output or action produced the observed error and recursively traces upstream, asking what earlier decision led to this intermediate failure, until either (i)~the preceding steps contain no error, or (ii)~the failure cause is unrelated to earlier trajectory decisions.
Each chain terminates at an \emph{error critical step}—the earliest decision that triggers the downstream cascade~\citep{letsverify2023,processbench2025,enconda2025}—and receives an \emph{error type label} from a controlled vocabulary (environment/setup issues, dependency resolution failures, mislocalized edits, incorrect hypotheses, verification misinterpretation, and unproductive looping).
To assess annotation reliability, a random 15\% subset of trajectories was independently double annotated; interannotator agreement on the error-critical step label reached Cohen's $\kappa = 0.73$.
The full annotation guidelines, including edge case handling, are provided in \Cref{sec:app:annotation}.

The annotation process surfaces several recurring patterns across backbones and agent frameworks that motivate the design of \system.

\begin{figure}[!t]
\centering
\includegraphics[width=\figCardTwoW]{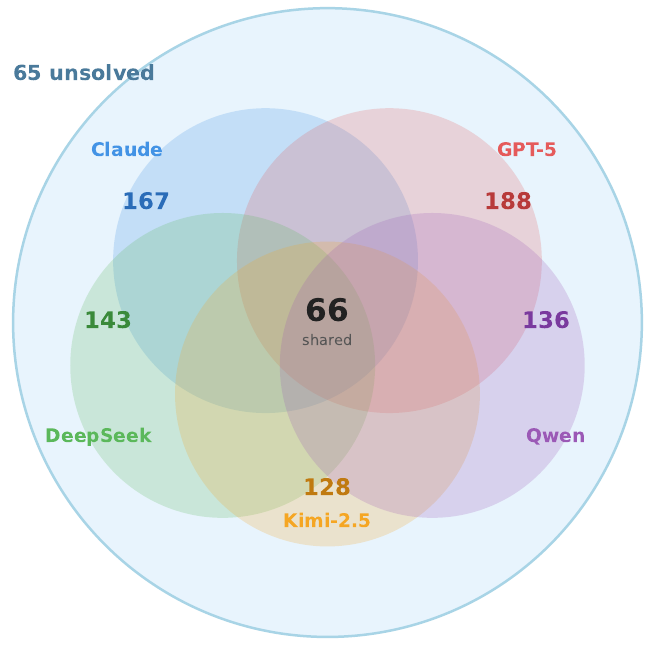}
\caption{Task categories solved per backbone. The central 66 categories are solved by all five models; 65 categories remain unsolved by any model.}
\label{fig:model_venn}
\vspace{-1mm}
\end{figure}

\enlargethispage{2\baselineskip}
\paragraph{Model preferences and shared failure modes.}
We analyze per-category pass rates across all 15 backbone--agent groups (\Cref{fig:model_venn}).
Among 340 task categories, the 66 categories solved by all five models are mostly routine transformation and scripting workloads, such as regex manipulation, JSON/CSV processing, and standard numerical procedures.
At the other extreme, the 65 categories unsolved by every model skew toward tasks that require harder external grounding or longer horizon reasoning, including formal verification, computer vision, advanced scientific computing, and legacy environments.
Between these two poles, the strongest backbones remain broadly competitive but exhibit clear task preferences: GPT-5 is relatively stronger on graph heavy, chemistry, and forensics tasks; Claude-sonnet-4 on Bayesian inference and speculative decoding; Kimi-K2-Instruct on graphics and ray tracing; and DeepSeek-V3.2 on data pipeline and package management workloads.
When none of the models can truly solve the task, their behaviors become strikingly similar: instead of explicitly recognizing failure, they often bypass the bottleneck through fabricated evidence, placeholder outputs presented as real results, or premature stopping after becoming trapped in unproductive loops.

\begin{takeawaybox}{Takeaway 1 for Task Preferences and Failure Masking Behaviors}
Backbones differ more in \emph{which} task regimes they solve reliably than in aggregate solved count, but on universally hard categories they converge to the same failure pattern: bypassing the problem through fabrication, placeholder outputs, or early stopping (\Cref{fig:model_venn}).
\end{takeawaybox}

\begin{table}[t]
\centering
\caption{End-to-end success versus process cost on the no intersection setting. Tok.(k) is the average token consumption per task in thousands. SWE-Agent and OpenHands consume 2$\times$ the tokens of MiniSWE-Agent for only modest success gains.}
\label{tab:arch_cost_success}
\begin{tabular}{lcccc}
\toprule
Agent & Succ. (\%) & Steps & Avg. Steps & Tok.(k) \\
\midrule
MiniSWE-Agent & 32.8 & 18,792 & 19.9 & 44.6 \\
Terminus 2    & 35.2 & 25,381 & 24.7 & 51.3 \\
SWE-Agent     & 37.5 & 43,216 & 41.2 & 86.7 \\
OpenHands     & 38.3 & 46,754 & 43.9 & 91.4 \\
\bottomrule
\end{tabular}
\end{table}

\paragraph{Overengineering: when complexity does not buy success.}
We compare a lightweight agent design (\textit{MiniSWE-Agent}) with progressively more complex frameworks (\textit{Terminus~2}, \textit{SWE-Agent}, \textit{OpenHands}) along two axes: end-to-end task success and \emph{process cost}, measured by both step count and token consumption (\Cref{tab:arch_cost_success}). Terminus~2 adds moderate orchestration overhead over MiniSWE-Agent (51.3k vs.\ 44.6k tokens), while SWE-Agent and OpenHands nearly double token usage (86.7k and 91.4k). Yet these additional expenditures yield only modest success gains (+2.4--5.5\,pp). This pattern suggests that, for general terminal-based coding tasks, success is primarily bounded by backbone capability, while architectural complexity often translates into higher cost and longer interaction loops rather than reliably better outcomes.

\begin{takeawaybox}{Takeaway 2 for Overengineering}
Additional orchestration increases cost without proportional success gains; backbone capability is the primary lever (\Cref{tab:arch_cost_success}).
\end{takeawaybox}

\begin{figure}[t]
    \centering
    \includegraphics[width=\figMainWide]{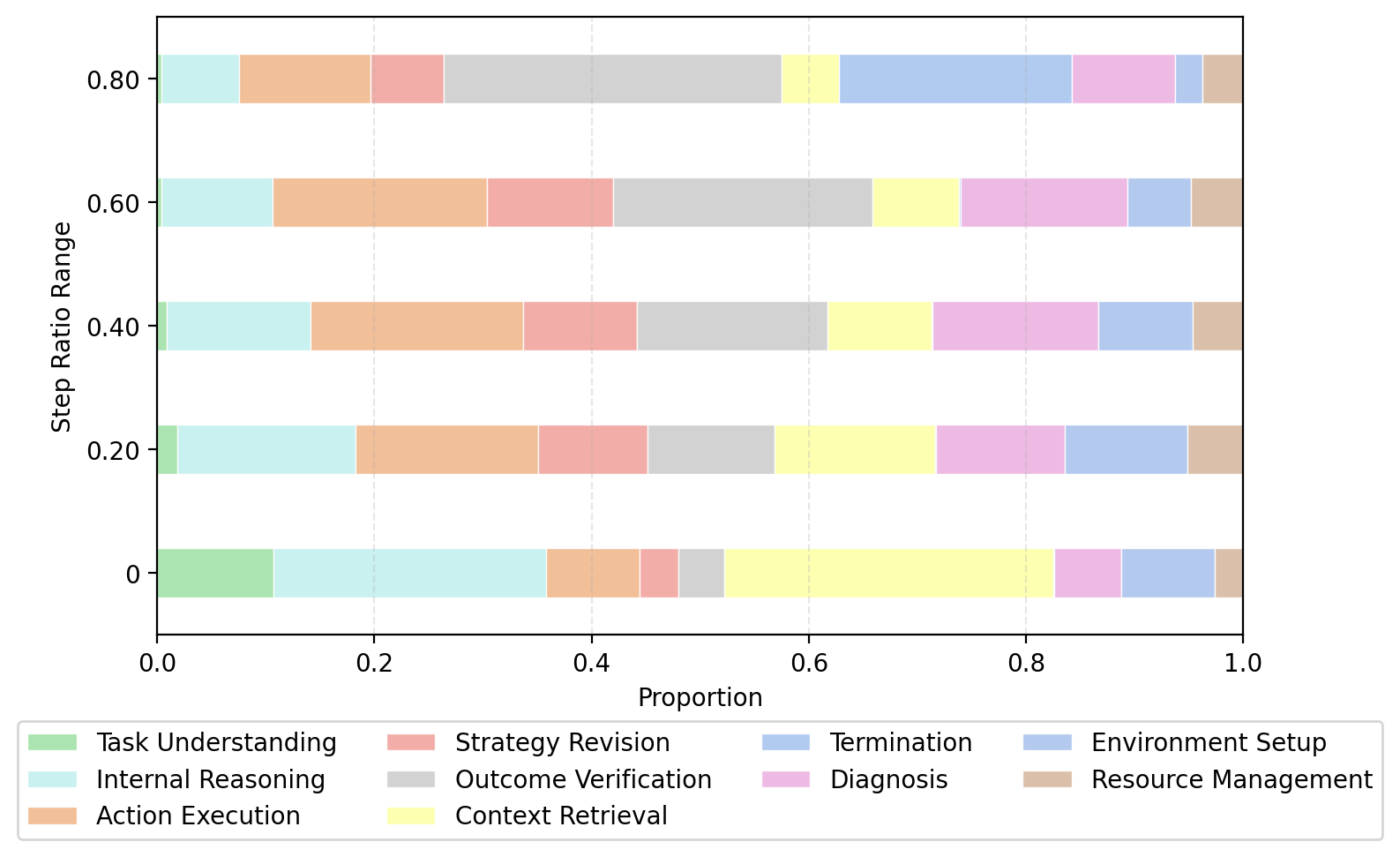}
    \caption{Distribution of error critical steps across stages, contrasting solved and unsolved runs.}
    \label{fig:error_stage_by_stage}
\end{figure}

\paragraph{Stage dynamics, failure onset, and error type shift.}
Using the stage labels assigned during annotation (environment verification, dependency installation, inspection/debugging, patching, and verification), we examine how stage occupancy and transitions relate to end-to-end success. Solved runs display a more coherent, forward moving progression with fewer back-and-forth oscillations. In contrast, unsolved runs allocate disproportionate budget to early setup activities and recurrent inspection loops, often after an early wrong commitment that later steps fail to undo. Error types also shift systematically across stages: environment and dependency errors concentrate early, while mislocalized edits, incorrect hypotheses, and verification misinterpretation dominate later patching and verification. The concentration of error critical steps in a small set of stages (\Cref{fig:error_stage_by_stage}) directly motivates our failure onset localization objective.

\begin{takeawaybox}{Takeaway 3 for Stage Dependent Errors}
Error type is largely determined by workflow phase: environment and dependency errors dominate early stages, while mislocalized edits and incorrect hypotheses concentrate in later patching and verification. This predictability enables stage-aware guardrails that can preempt failures before they cascade, rather than relying solely on post-hoc diagnosis (\Cref{fig:error_stage_by_stage}).
\end{takeawaybox}

\newcommand{\sat}[1]{\underline{#1}}
\newcommand{\plateau}[1]{\cellcolor{gray!15}#1}
\begin{table*}[t]
\centering
\caption{\textbf{Resolved rate (\%) and token consumption (in thousands) across iteration budgets and backbones}.
\texttt{Tok.(k)} reports average token usage within the first \texttt{max\_iter} rounds over all valid samples with extractable prefix token traces.
Token coverage by model and budget is reported separately for auditability.}
\label{tab:success_vs_iterations_3agent_merged}
\vspace{2mm}
\footnotesize
\setlength{\tabcolsep}{4.2pt}
\begin{tabular}{r cc cc cc cc cc}
\toprule
& \multicolumn{2}{c}{\textbf{Claude-sonnet-4}} & \multicolumn{2}{c}{\textbf{GPT-5}} &
\multicolumn{2}{c}{\textbf{DeepSeek-V3.2}} & \multicolumn{2}{c}{\textbf{Qwen3-Coder-480B}} & \multicolumn{2}{c}{\textbf{Kimi-K2-Instruct}} \\
\cmidrule(lr){2-3}\cmidrule(lr){4-5}\cmidrule(lr){6-7}\cmidrule(lr){8-9}\cmidrule(lr){10-11}
\texttt{max\_iter} & Res.\% & Tok.(k) & Res.\% & Tok.(k) & Res.\% & Tok.(k) & Res.\% & Tok.(k) & Res.\% & Tok.(k) \\
\midrule
10 & 12.67 & 116.98 & 38.69 & 106.19 & 12.43 & 86.79 & 9.76 & 99.38 & 10.61 & 78.81 \\
20 & 30.13 & 266.53 & 45.48 & 163.89 & 27.50 & 212.60 & 18.22 & 255.66 & 23.02 & 192.28 \\
40 & 38.00 & 493.46 & \plateau{47.06} & 214.60 & 38.18 & 400.96 & 28.85 & 543.64 & 29.14 & 402.20 \\
100 & \plateau{41.65} & 767.77 & \plateau{47.06} & 249.80 & \plateau{39.75} & 479.77 & 31.67 & 970.46 & 32.55 & 715.24 \\
$\geq$150 & \plateau{41.65} & 886.00 & \plateau{47.06} & 266.32 & \plateau{39.75} & 512.42 & 32.10 & 1182.06 & 33.27 & 805.38 \\
\bottomrule
\end{tabular}
\end{table*}

\paragraph{Success saturates with additional iterations.}
We sweep \texttt{max\_iterations} over $\{5,\dots,300\}$ and reevaluate every backbone--framework pair under each budget. \Cref{tab:success_vs_iterations_3agent_merged} reports the resolved rates and token usage aggregated at the backbone level, while the full $5\times 3$ per combination trajectories are provided in \Cref{fig:iter_grid_app}. Two consistent patterns emerge. First, success improves rapidly as the iteration budget increases to around 40 steps, but the gains quickly diminish thereafter and the curves largely flatten. Second, the \emph{saturation ceiling} is mainly backbone dependent: stronger models plateau at higher resolved rates, but not substantially later. This suggests that extra iterations mainly help recover from low budget underexploration, rather than fundamentally improving the agent’s reasoning ability. When an agent commits early to an incorrect hypothesis, additional iterations are often spent on redundant exploration and trial-and-error edits rather than correcting the underlying mistake.

\begin{takeawaybox}{Takeaway 4 for Iteration Budget}
Success saturates quickly; extra iterations amplify unproductive loops rather than enabling recovery (\Cref{tab:success_vs_iterations_3agent_merged}).
\end{takeawaybox}

\FloatBarrier
\section{CodeTracer}
\label{sec:tracing}

\subsection{Method}
\label{sec:ta_tracer_system}

\system tackles \textbf{failure onset localization}: given a standardized trajectory $\tau$ segmented into ordered stages, it predicts a failure responsible stage $\hat{s}$, a set of error relevant steps $P$ within $\hat{s}$, and a compact evidence set $E$ supporting the diagnosis.
Supervision marks stages that are causally responsible for the eventual failure under our annotation guidelines, and evaluation measures step level Precision/Recall/F1 against the gold incorrect step set $G$ plus token cost under matched budgets.

Because run directories are heterogeneous and their file layouts are agent dependent, hardcoded parsers are brittle. \system therefore separates \emph{evolving extraction}, \emph{tree indexing}, and \emph{diagnosis}, while accumulating reusable parsers so that support for new formats improves over time, following the broader need for process-aware trajectory analysis in software engineering agents~\citep{bouzenia2025trajectories,enconda2025}.

\begin{figure}[t]
\centering
\includegraphics[width=1.0\linewidth]{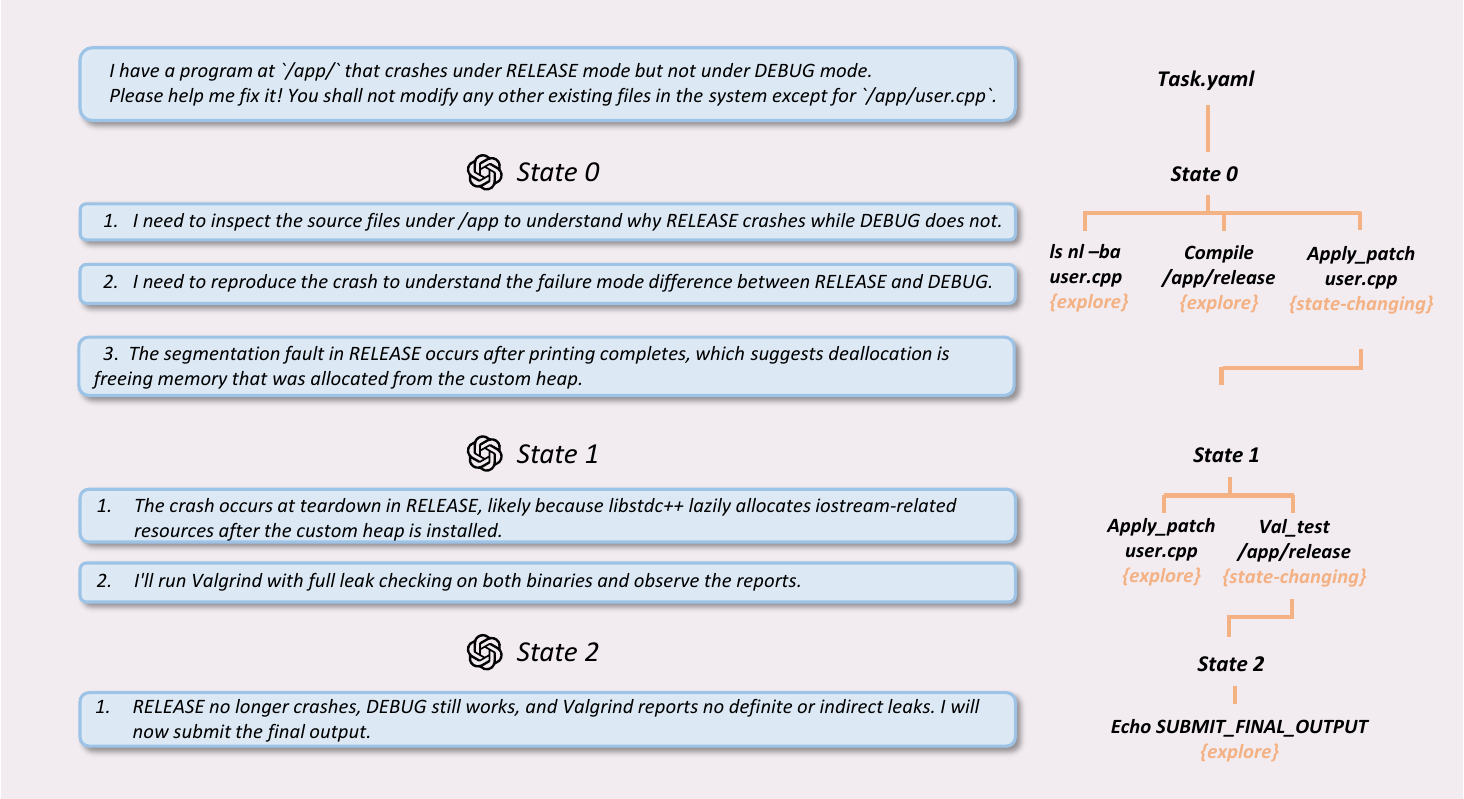}
\caption{Hierarchical trace tree. Exploration steps remain under the current state node, whereas state changing steps induce transitions to child states. \system builds this structure during tree indexing to provide compressed navigation of execution traces.}
\label{fig:trace_format_example}
\end{figure}

\parabf{Pipeline.}
\system decomposes tracing into three stages (details in \Cref{sec:app:framework}):
(1)~\textbf{Evolving extraction} scans a run directory, produces a compact layout spec describing which artifacts record execution steps, and adapts a parser accordingly: it checks a registry of existing parsers; if no match is found, it synthesizes and registers a new one, then instantiates a format-specific diagnosis prompt with a fixed query schema. This stage emits normalized step records with typed fields (action, observation, diff, verification outcome) while preventing format drift across runs.
(2)~\textbf{Tree indexing} converts the flat sequence of normalized steps into a \emph{hierarchically structured} trace tree (\Cref{fig:trace_format_example}). Steps that only inspect the current environment without modifying the codebase or execution state are represented as exploration nodes under the same state, whereas steps that modify the codebase or execution environment are represented as state-changing nodes that induce transitions to child states. Each node is further annotated with a summary of intent and outcome. The resulting tree makes explicit whether subsequent actions are taken under an unchanged context or after an intervention, and serves as a compressed navigation index for diagnosis.
(3)~\textbf{Diagnosis} traverses the tree, issues structured evidence queries, and outputs the failure responsible stage, error-relevant steps, and a compact evidence set justifying the localization.

\subsection{CodeTraceBench}
\label{sec:tracebench}

We construct \bench directly from the trajectory corpus in \Cref{sec:study}. The source pool spans five benchmarks (SWE-bench Verified, SWE-bench Pro, MultiSWE-bench, SWE-PolyBench, and TerminalBench), five backbones (Claude-sonnet-4, GPT-5, Kimi-K2-Instruct, Qwen3-Coder-480B, DeepSeek-V3.2), and four frameworks (SWE-Agent, MiniSWE-Agent, OpenHands, Terminus~2), totaling 4{,}354 standardized and step-level annotated trajectories~\citep{ICLR2024_edac78c3,NEURIPS2024_5a7c9475,ICLR2025_a4b6ad6b}. From this pool, we curate long horizon instances with clear failure cascades and sufficient in-trace evidence, while removing short or near-duplicate runs.

The resulting benchmark is produced with a \texttt{full} split (3.32K instances) and a higher quality \texttt{verified} split (1.06K). Each instance records the framework, backbone, task metadata (236 tasks, 26 categories, difficulty labels), raw artifact pointers, stage boundaries, failure critical stage labels, and incorrect step annotations, supporting stage localization and within stage evidence retrieval.

\begin{table*}[t]
\centering
\caption{\textbf{Main localization results on CodeTraceBench (macro, step level) stratified by difficulty.}
P/R/F1 are reported as percentages (\%).
Tok denotes total LLM tokens (in thousands).
Overall uses the full evaluation set; Easy/Medium/Hard split by trajectory difficulty.}
\label{tab:tracebench_main}
\vspace{2mm}
\footnotesize
\resizebox{\textwidth}{!}{%
\begin{tabular}{l cccc cccc cccc cccc}
\toprule
& \multicolumn{4}{c}{\textbf{Overall}} & \multicolumn{4}{c}{\textbf{Easy}} & \multicolumn{4}{c}{\textbf{Medium}} & \multicolumn{4}{c}{\textbf{Hard}} \\
\cmidrule(lr){2-5}\cmidrule(lr){6-9}\cmidrule(lr){10-13}\cmidrule(lr){14-17}
Backbone & P & R & F1 & Tok
         & P & R & F1 & Tok
         & P & R & F1 & Tok
         & P & R & F1 & Tok \\
\midrule
\multicolumn{17}{l}{\textbf{Bare LLM}} \\
\textsc{Claude-sonnet-4}    & 16.64 & 15.82 & 16.22 & 105.1 & 18.31 & 17.08 & 17.67 & 68.5  & 16.07 & 15.14 & 15.59 & 97.8  & 13.52 & 12.83 & 13.17 & 155.6 \\
\textsc{GPT-5}         & 16.69 & 21.46 & 18.78 & 58.5  & 18.42 & 23.54 & 20.65 & 38.2  & 16.14 & 20.73 & 18.13 & 54.5  & 16.94 & 17.42 & 17.18 & 86.6  \\
\textsc{DeepSeek-V3.2} & 13.11 & 21.66 & 16.33 & 83.4  & 14.37 & 23.72 & 17.87 & 54.6  & 12.58 & 20.74 & 15.67 & 77.5  & 10.28 & 17.95 & 12.89 & 123.6 \\
\midrule
\multicolumn{17}{l}{\textbf{Mini-CodeTracer}} \\
\textsc{Claude-sonnet-4}    & 20.50 & 26.24 & 19.17 & 82.4 & 23.58 & 28.46 & 22.04 & 54.2 & 19.83 & 29.17 & 18.42 & 77.1 & 15.47 & 20.83 & 14.08 & 120.6 \\
\textsc{GPT-5}         & 26.03 & 21.39 & 19.33 & 44.8 & 28.71 & 24.26 & 22.31 & 29.5 & 25.42 & 20.63 & 18.68 & 42.0 & 20.73 & 16.18 & 14.17 & 65.8  \\
\textsc{DeepSeek-V3.2} & 24.08 & 20.98 & 19.24 & 63.8 & 27.14 & 23.62 & 21.68 & 42.0 & 23.24 & 19.87 & 18.43 & 59.5 & 18.27 & 16.41 & 14.62 & 94.2  \\
\midrule
\multicolumn{17}{l}{\textbf{CodeTracer}} \\
\textsc{Claude-sonnet-4}    & 40.47 & \textbf{\underline{54.87}} & 46.57 & 56.8 & 45.28 & \textbf{\underline{59.13}} & 51.29 & 37.6 & 39.36 & \textbf{\underline{53.58}} & 45.38 & 53.4 & 32.71 & \textbf{\underline{47.24}} & 38.67 & 82.5  \\
\textsc{GPT-5}         & \textbf{\underline{45.02}} & 51.46 & \textbf{\underline{48.02}} & \textbf{\underline{31.1}} & \textbf{\underline{49.86}} & 55.78 & \textbf{\underline{52.68}} & \textbf{\underline{20.6}} & \textbf{\underline{43.81}} & 50.42 & \textbf{\underline{46.84}} & \textbf{\underline{29.2}} & \textbf{\underline{37.08}} & 43.79 & \textbf{\underline{40.14}} & \textbf{\underline{45.7}} \\
\textsc{DeepSeek-V3.2} & 43.17 & 49.58 & 46.14 & 44.6 & 47.68 & 53.94 & 50.62 & 29.5 & 42.02 & 48.53 & 45.04 & 41.8 & 35.74 & 42.21 & 38.72 & 65.3  \\
\bottomrule
\end{tabular}}
\end{table*}

\section{Experiments}
\label{sec:exp}

\subsection{Setup}
\label{sec:exp:setup}

We evaluate  CodeTraceBench using the intersection subset (backbone--method pairs with complete coverage).
We test Claude-sonnet-4, GPT-5, and DeepSeek-V3.2 under three localization methods (Bare LLM, \miniSystem, \system), all operating on the same run directories under matched budgets and unified decoding settings.
We report macro averaged step level Precision, Recall, and F1 (formal definitions in \Cref{sec:app:metrics}) and total token usage.

\subsection{Main Results}
\label{sec:exp:main}

\Cref{tab:tracebench_main} summarizes the main localization results. Across all backbones, adding structure and tracing signals substantially improves localization quality over raw log prompting. \miniSystem already recovers part of the gain, showing that lightweight standardization is a strong baseline, while \system further improves both precision and recall by prioritizing genuinely failure relevant steps rather than merely salient artifacts. Token efficiency is also improved because tracing narrows evidence retrieval to compact candidate sets.
The three frontier backbones reach comparable F1 (46--48\%) but diverge in how they traverse the trace.
GPT-5 terminates its search earlier, committing to a compact set of failure relevant steps with high confidence; this yields the best precision (45.0\% overall, 49.9\% Easy) and the lowest token cost (31.1k overall, 20.6k Easy) but misses some error contributing steps.
Claude-sonnet-4 continues scanning deeper into the trace before concluding, surfacing more failure relevant evidence and achieving the highest recall across all splits (54.9\% overall, 59.1\% Easy) at the expense of higher token consumption (56.8k) and lower precision (40.5\%).
DeepSeek-V3.2 falls between the two, balancing coverage and conciseness with the most uniform P/R gap across difficulty levels.
Harder tasks correlate with longer trajectories and proportionally higher token budgets---the Easy to Hard token ratio roughly doubles for every backbone---confirming that localization difficulty scales with trajectory complexity.
We further stratify results by difficulty, category, and execution stage (Appendix); the relative ordering among methods remains stable, with \system providing the largest gains on harder instances where failure relevant evidence is dispersed.

\begin{takeawaybox}{Takeaway 5 for Diagnosis Behaviors}
Frontier backbones achieve similar F1 but exhibit different search depth strategies: GPT-5 commits early for higher precision at lower cost, Claude-sonnet-4 searches exhaustively for higher recall, and DeepSeek-V3.2 balances both. Localization cost scales with trajectory length, not model identity (\Cref{tab:tracebench_main}).
\end{takeawaybox}

\subsection{Analysis and Replay}
\label{sec:exp:integration}

We conduct analyses using \system's hierarchical trace structure and diagnostic outputs.

\begin{figure}[t]
    \centering
    \includegraphics[width=\figMainMedium]{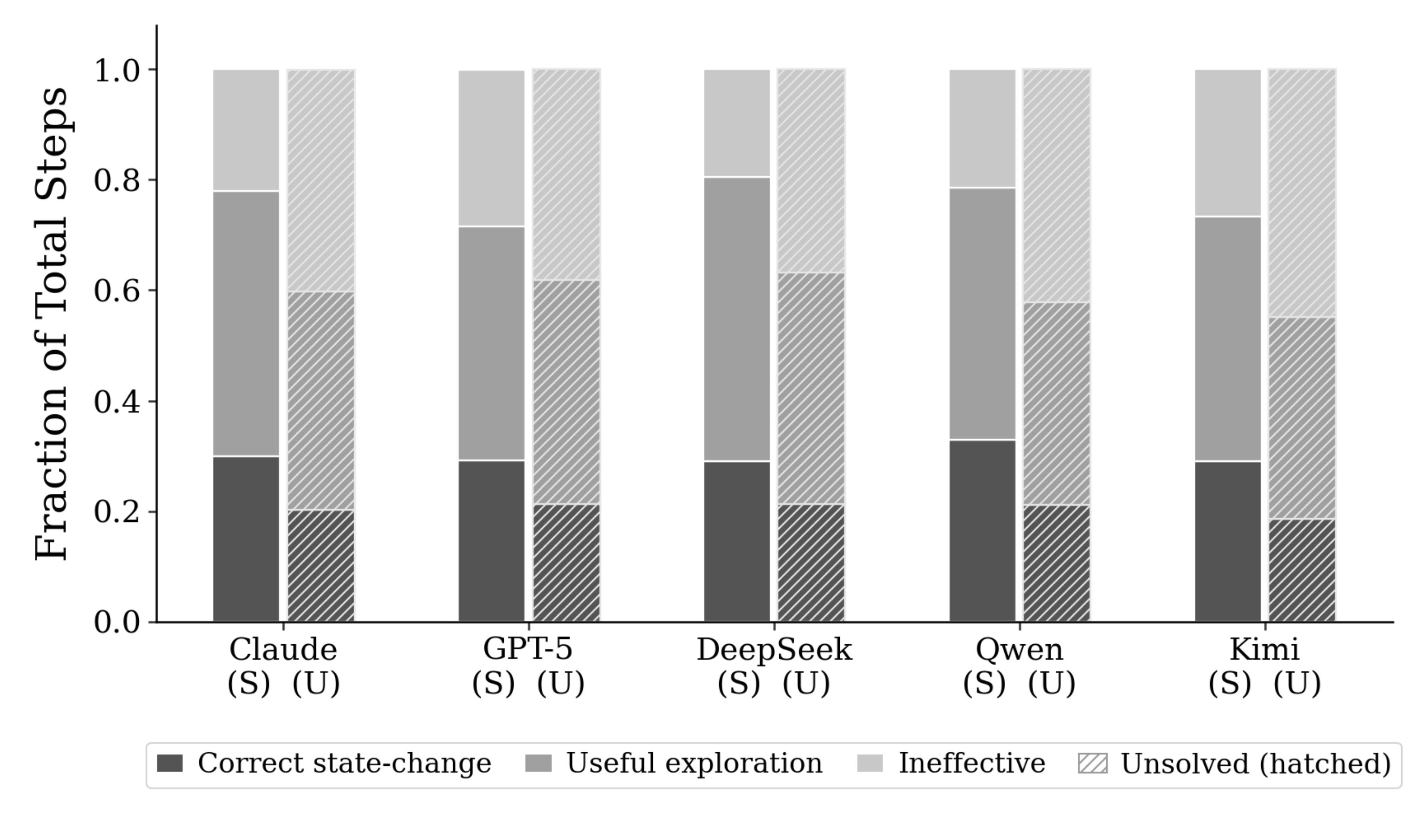}
    \caption{Step budget decomposition per backbone (solved vs.\ unsolved) on the intersection subset. Each bar shows the fraction of total steps that are correct state changes, useful exploration, or ineffective. Hatched bars denote unsolved trajectories.}
    \label{fig:evidence_action_gap}
\end{figure}

\paragraph{Evidence to action gap.}
\Cref{fig:evidence_action_gap} decomposes each trajectory's step budget into correct state changes, useful exploration, and ineffective steps.
Across all five models the ineffective fraction nearly doubles from solved (22\%) to unsolved (40\%), while correct state changes drop consistently (30\%$\to$21\%).
Exploration usefulness degrades only mildly, indicating that agents still gather relevant information but fail to translate it into correct actions---a comprehension bottleneck whose severity varies by backbone (Qwen3-Coder-480B and Kimi-K2-Instruct show the sharpest drops, $\Delta\!=\!$11.7 and 10.3\,pp).

\begin{takeawaybox}{Takeaway 6 for Evidence to Action Gap}
Agents often gather useful evidence through exploration, but fail to translate it into effective state changing actions (\Cref{fig:evidence_action_gap}).
\end{takeawaybox}

\begin{table*}[t]
\centering
\caption{\textbf{Component ablation.}
Each row incrementally adds one component on top of \miniSystem (diagnosis loop).
Evolving extraction adds layout discovery and parser adaptation with accumulating reuse;
tree indexing builds a hierarchical trace tree for compressed navigation.
P/R/F1 are step level macro metrics (\%); Tok is average tokens per instance (k).}
\label{tab:ablation}
\vspace{2mm}
\footnotesize
\resizebox{0.9\textwidth}{!}{%
\begin{tabular}{l cccc cccc cccc}
\toprule
& \multicolumn{4}{c}{\textbf{\textsc{Claude-sonnet-4}}} & \multicolumn{4}{c}{\textbf{\textsc{GPT-5}}} & \multicolumn{4}{c}{\textbf{\textsc{DeepSeek-V3.2}}} \\
\cmidrule(lr){2-5}\cmidrule(lr){6-9}\cmidrule(lr){10-13}
Configuration & P & R & F1 & Tok & P & R & F1 & Tok & P & R & F1 & Tok \\
\midrule
Bare LLM                & 16.64 & 15.82 & 16.22 & 105.1 & 16.69 & 21.46 & 18.78 & 58.5 & 13.11 & 21.66 & 16.33 & 83.4 \\
\miniSystem             & 20.50 & 26.24 & 19.17 & 82.4  & 26.03 & 21.39 & 19.33 & 44.8 & 24.08 & 20.98 & 19.24 & 63.8 \\
+ Evolving Extraction   & 27.83 & 33.45 & 28.12 & 71.0  & 31.56 & 29.87 & 29.45 & 37.9 & 29.74 & 28.91 & 28.38 & 55.3 \\
+ Tree Indexing   & \textbf{\underline{40.47}} & \textbf{\underline{54.87}} & \textbf{\underline{46.57}} & \textbf{\underline{56.8}}  & \textbf{\underline{45.02}} & \textbf{\underline{51.46}} & \textbf{\underline{48.02}} & \textbf{\underline{31.1}} & \textbf{\underline{43.17}} & \textbf{\underline{49.58}} & \textbf{\underline{46.14}} & \textbf{\underline{44.6}} \\
\bottomrule
\end{tabular}}
\end{table*}

\paragraph{Component ablation.}
We incrementally add \system's two key components---evolving extraction (layout discovery + parser adaptation) and tree indexing---on top of \miniSystem (\Cref{tab:ablation}).
Tree indexing drives the largest single gain (18.3\,pt F1) by providing compressed hierarchical navigation, while evolving extraction contributes a further 9.4-point lift through format standardization with parser reuse.

\begin{figure}[t]
\centering
\includegraphics[width=\figMainMedium]{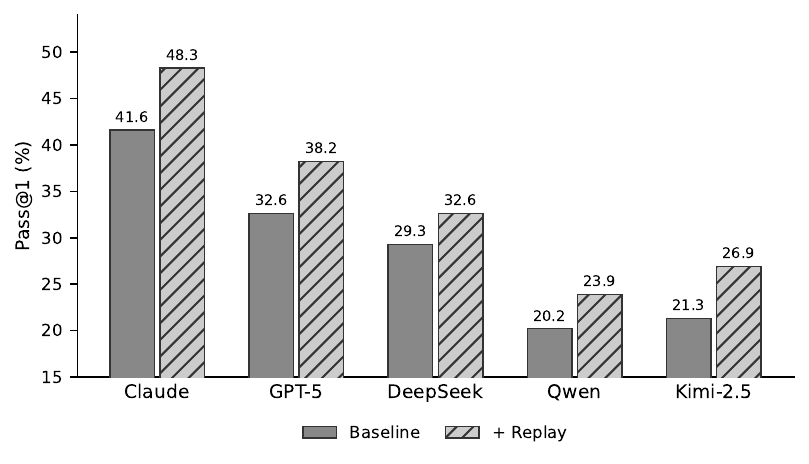}
\caption{\textbf{Reflective replay.} Pass@1 on originally failed runs before and after injecting \system's diagnostic signals under matched budgets.}
\label{fig:replay}
\end{figure}

\paragraph{Reflective replay.}
We feed \system's localized evidence back into agents: on originally failed runs, the same backbone is reinvoked under matched budget with the diagnosis injected as a prefix hint (\Cref{fig:replay})~\citep{shinn2023reflexion,madaan2023selfrefine,chen2024selfdebug}.
Replay consistently improves Pass@1 across all backbones, helping agents revise early wrong commitments instead of repeating unproductive exploration.
The diagnosis pass itself consumes on average 8.4k tokens for Claude-sonnet-4, 5.2k for GPT-5, and 7.1k for DeepSeek-V3.2; these tokens are counted \emph{outside} the replay budget so that the replayed run receives exactly the same iteration and token budget as the original.

\paragraph{Industrial agent analysis.}
Beyond the academic agent frameworks studied above, we apply \system to analyze trajectories from Claude Code, an industrial coding agent. Compared to research agents, Claude Code employs substantially richer tooling infrastructure (40+ specialized tools across 8 categories) and sophisticated context management (compaction, token budgeting), yielding a lower exploration-to-change ratio that correlates with higher trajectory efficiency; however, parallel tool execution---unique to industrial agents---introduces ordering-sensitivity issues absent from sequential academic frameworks. Full architectural comparison and trajectory statistics are provided in \Cref{sec:app:claude_code}.

\begin{figure*}[t]
  \centering
  \begin{subfigure}[t]{0.30\linewidth}
    \centering
    \includegraphics[width=\linewidth,trim=0 0 0 15,clip]{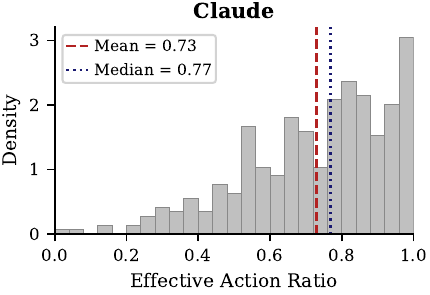}
    \caption{Claude-sonnet-4}
  \end{subfigure}\hfill
  \begin{subfigure}[t]{0.30\linewidth}
    \centering
    \includegraphics[width=\linewidth,trim=0 0 0 15,clip]{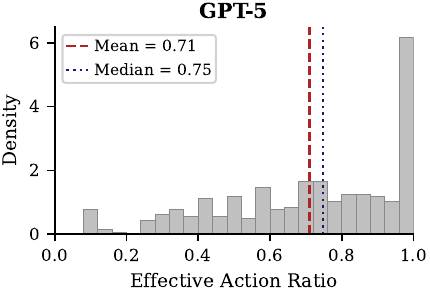}
    \caption{GPT-5}
  \end{subfigure}\hfill
  \begin{subfigure}[t]{0.30\linewidth}
    \centering
    \includegraphics[width=\linewidth,trim=0 0 0 12.5,clip]{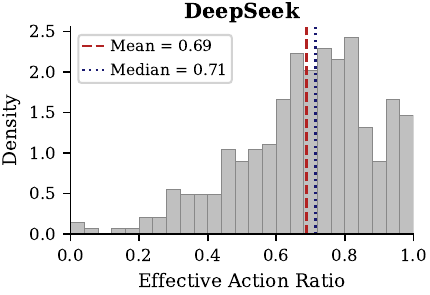}
    \caption{DeepSeek-V3.2}
  \end{subfigure}

  \vspace{1mm}
  \begin{subfigure}[t]{0.30\linewidth}
    \centering
    \includegraphics[width=\linewidth,trim=0 0 0 15,clip]{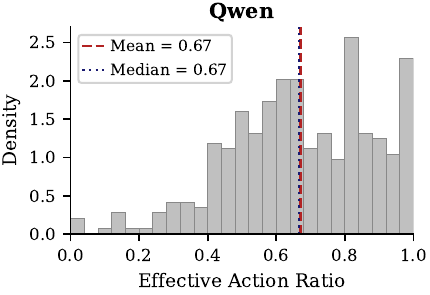}
    \caption{Qwen3-Coder-480B}
  \end{subfigure}\hfill
  \begin{subfigure}[t]{0.30\linewidth}
    \centering
    \includegraphics[width=\linewidth,trim=0 0 0 15,clip]{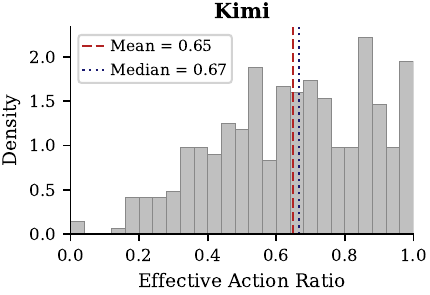}
    \caption{Kimi-K2-Instruct}
  \end{subfigure}\hfill
  \begin{subfigure}[t]{0.30\linewidth}
    \centering
    \includegraphics[width=\linewidth]{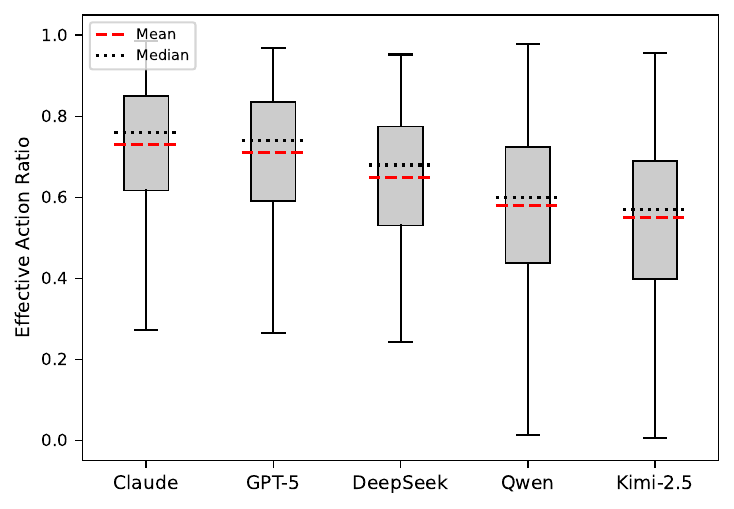}
    \caption{Cross model summary}
  \end{subfigure}
  \caption{\textbf{Effective action ratio.} (a)--(e)~per model histograms; (f)~cross model violin and box summary.}
  \label{fig:effective_action_ratio}
\end{figure*}

\paragraph{Action efficiency.}
We define the \emph{effective action ratio} as the fraction of goal-advancing steps per trajectory.
\Cref{fig:effective_action_ratio} shows that stronger models achieve higher means (73\% for Claude-sonnet-4, 71\% for GPT-5), yet all exhibit broad distributions with pronounced left tails---every backbone contains a nontrivial share of trajectories where fewer than half of actions are goal advancing.

\begin{takeawaybox}{Takeaway 7 for Action Efficiency}
Even strong models execute many ineffective actions; improving early unproductive detection matters as much as adding steps (\Cref{fig:effective_action_ratio}).
\end{takeawaybox}

\section{Related Work}
\label{sec:rw}



\paragraph{Code Agents and Software Engineering Benchmarks.}
Recent work has advanced software engineering agents in realistic development settings, including repository-level bug fixing and long-horizon terminal interaction. Representative systems and paradigms include SWE-Agent, OpenHands, CodeAct-style executable action agents, and Agentless, while evaluation has increasingly centered on executable benchmarks such as SWE-bench and Terminal-Bench~\citep{NEURIPS2024_5a7c9475,ICLR2025_a4b6ad6b,pmlr-v235-wang24h,xia2025demystifying,ICLR2024_edac78c3,merrill2026terminalbench}. However, these works primarily emphasize end-to-end task success, patch correctness, and resolved rates, offering limited visibility into where a long agent trajectory first becomes failure critical.

\paragraph{Trajectory Error Localization, Debugging, and Replay.}
Recent work has also begun to study software engineering agent trajectories and process-level quality~\citep{bouzenia2025trajectories,enconda2025}. Related benchmark efforts in other domains further show that step-level supervision and process error identification can expose failures that outcome only evaluation misses~\citep{letsverify2023,processbench2025}. Our localization objective is also related to issue localization, classical fault localization, and LLM-based debugging: prior work has localized issue relevant code regions for software engineering tasks, ranked suspicious program elements from failing executions, and leveraged runtime traces to debug candidate programs~\citep{xia2025demystifying,jones2005tarantula,abreu2009sfl,zhong2024debuglikehuman,chen2024selfdebug}. Reflection based self-improvement methods such as Reflexion and Self-Refine further show how critique, feedback, and memory can improve subsequent attempts~\citep{shinn2023reflexion,madaan2023selfrefine}. In contrast, we focus on hierarchical trace standardization, stage level failure onset localization, evidence retrieval, and replay oriented diagnostic outputs on executed agent trajectories.

For completeness and reproducibility, the appendix summarizes the full annotation protocol, additional framework and metric details, and the complete prompts and output schemas used for diagnosis and replay, together with extended empirical results, including per backbone and per framework breakdowns across difficulty settings and iteration budgets, that support the main text findings.

\section{Conclusion}
\label{sec:conclusion}

We presented \textsc{CodeTracer}, a tracing framework that converts heterogeneous code agent logs into structured hierarchical traces and performs automated failure onset localization, together with \textsc{CodeTraceBench}, a benchmark of thousands of step level annotated trajectories spanning four agent frameworks, five frontier backbones, and diverse software engineering tasks. Our empirical study across this corpus revealed that agents frequently gather the right diagnostic evidence yet fail to act on it, that extra orchestration complexity and iteration budget yield diminishing returns once the backbone reasoning ceiling is reached, and that error types shift predictably across workflow stages. \textsc{CodeTracer} substantially outperforms raw log prompting and lightweight baselines in step level localization, reaching up to 48\% macro F1 with lower token cost, and injecting its diagnostic outputs through reflective replay consistently recovers originally failed runs under matched budgets.

\newpage
\section*{Limitations}
Our study has several limitations. First, although CodeTraceBench is constructed from executed trajectories spanning multiple agent frameworks, backbones, and workloads, it does not cover the full design space of software engineering agents or real world repositories; conclusions about failure patterns and replay gains may therefore not transfer uniformly to other frameworks, domains, or deployment settings. Second, our supervision relies on stage and step level annotation of failure critical behavior. While our guidelines aim to make these judgments consistent, labels such as ``incorrect,'' ``unuseful,'' and ``error critical'' still involve annotator interpretation, especially for long trajectories with intertwined exploration and state changes. Third, our tracing and replay results are evaluated under matched offline budgets on previously collected runs. This setting isolates diagnostic quality, but it cannot fully capture how agents might adapt to interactive human oversight, changing environments, or repeated online intervention. Finally, our reflective replay experiments test whether localized evidence can help recover failed runs, but they do not by themselves establish a general training signal or guarantee robust improvement across all task categories and model families. We therefore view \system and \bench as tools for structured diagnosis and controlled evaluation, rather than as a complete account of code agent reliability.

\bibliographystyle{plainnat}
\bibliography{references}

\appendix
\onecolumn

\section{Annotation Guidelines}
\label{sec:app:annotation}

This appendix provides detailed definitions for the annotation schema described in \Cref{sec:annotation}. Each annotator receives the task specification, the reference solution, and access to the execution environment, and is assigned all 15 backbone--agent trajectories for each task.

\parabf{Stage labeling (all trajectories).}
Every step in a trajectory is assigned a \emph{stage label} from the following ordered vocabulary:
(i)~\textit{environment verification} — confirming the runtime, toolchain, and test harness;
(ii)~\textit{dependency installation} — installing, upgrading, or pinning packages and system libraries;
(iii)~\textit{inspection / debugging} — reading source files, searching code, running tests, examining logs;
(iv)~\textit{patching} — modifying source code, configuration files, or build scripts;
(v)~\textit{verification} — rerunning tests, linting, or other validation to confirm the patch.
A trajectory may revisit stages (e.g., returning to inspection after a failed patch); in such cases each contiguous block receives its own stage span.

\parabf{Successful trajectories.}
For trajectories that pass all tests, the annotator identifies two categories of noncontributing steps:
(i)~\emph{Redundant} steps: actions whose effects are fully subsumed by earlier steps (e.g., rereading a file already inspected, reapplying an identical patch).
(ii)~\emph{Trial-and-error} steps: actions that are later reverted or superseded (e.g., an incorrect patch attempt that is subsequently overwritten by the correct fix).

\parabf{Failed trajectories — chain based backward tracing.}
For trajectories that fail, the annotator performs recursive backward tracing from the failing test output:
(i)~identify the immediately preceding step whose output or action produced the observed error;
(ii)~trace upstream by asking which earlier decision led to this intermediate failure;
(iii)~repeat until the preceding steps contain no error or the failure cause is unrelated to earlier trajectory decisions.
The chain terminates at an \emph{error critical step} — the earliest decision that triggers the downstream cascade.
Each error critical step receives an \emph{error type label} from a controlled vocabulary:
environment/setup issues, dependency resolution failures, mislocalized edits, incorrect hypotheses, verification misinterpretation, and unproductive looping.

\parabf{Edge cases.}
(i)~If a step is ambiguous (e.g., a partially correct patch), annotators default to \emph{incorrect} and note the ambiguity.
(ii)~Environment setup steps that fail due to external factors (network timeouts, missing packages) are labeled \emph{incorrect} only if the agent chose an avoidable path; annotators verify by entering the execution environment.
(iii)~When multiple stages contain incorrect steps, the \emph{earliest} stage with error critical steps is designated as the failure responsible stage.



\begin{figure}[htbp]
    \centering
    \begin{minipage}[t]{0.28\textwidth}
        \centering
        \includegraphics[width=\textwidth]{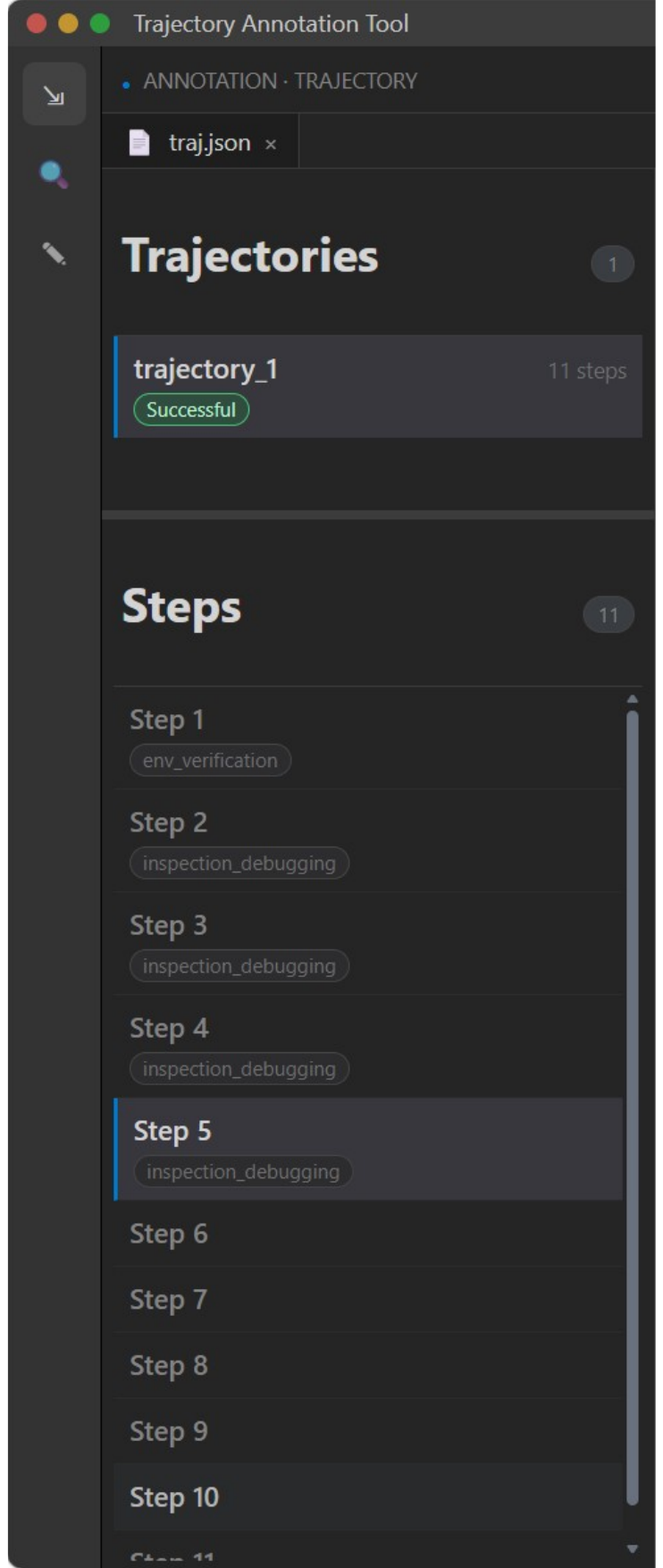}
        \caption{Annotation tool interface — sidebar components}
        \label{fig:annotation-sidebar}
    \end{minipage}
    \hfill
    \begin{minipage}[t]{0.68\textwidth}
        \centering
        \includegraphics[width=\textwidth]{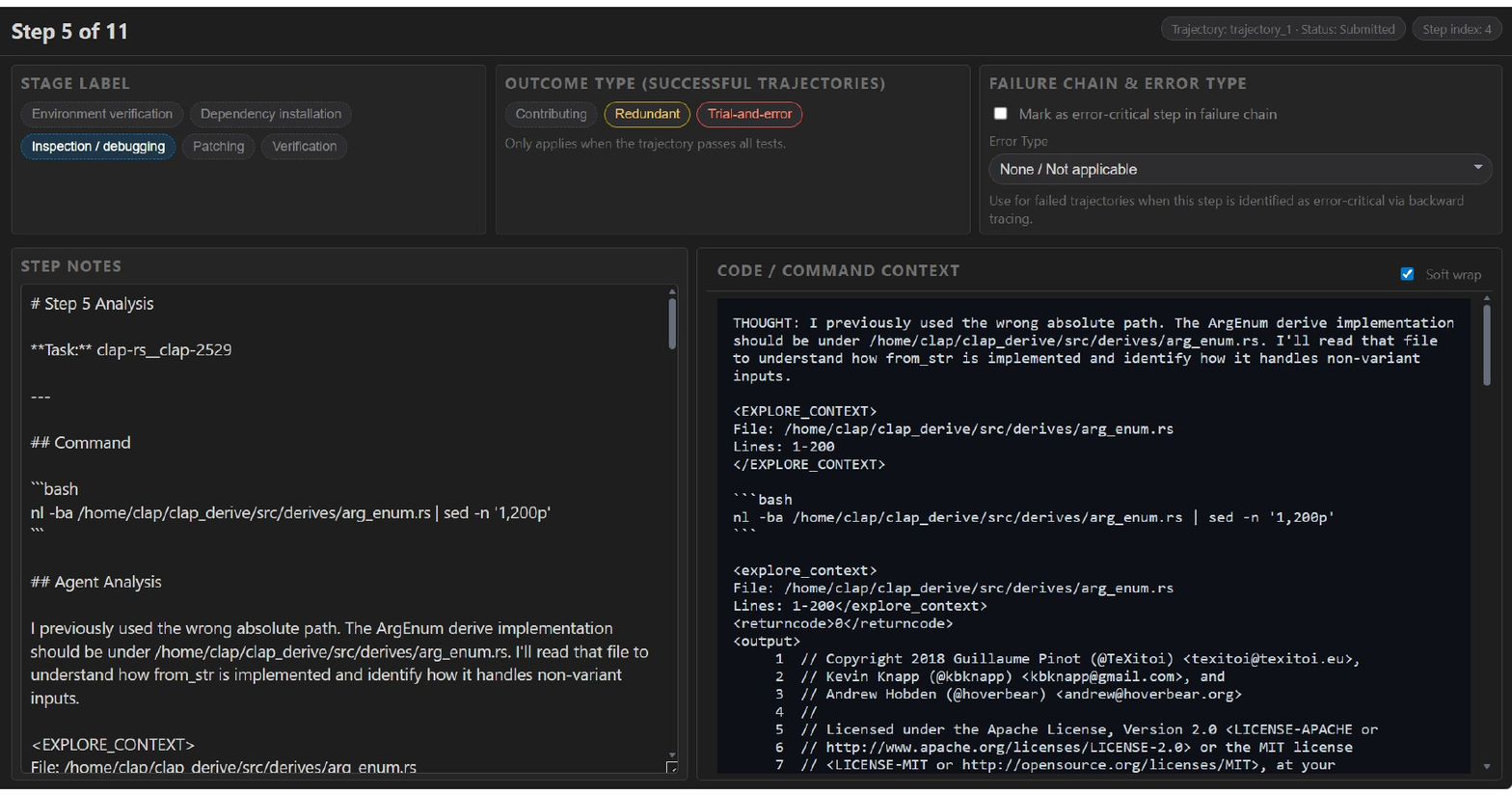}
        \caption{Annotation tool interface — main panel}
        \label{fig:annotation-main}
    \end{minipage}
\end{figure}

\section{Tracing Framework Details}
\label{sec:app:framework}

The three-stage pipeline of \system is summarized in \Cref{sec:ta_tracer_system}. Below we provide additional details on the scoring features, replay protocol, and computational complexity.

\parabf{Scoring features.}
The trace agent uses the following signals for ranking candidate stages:
(i)~\textit{verification regression}: whether a stage's state changing steps cause a previously passing test to fail;
(ii)~\textit{diff magnitude}: cumulative lines changed in state changing steps within the stage;
(iii)~\textit{backtrack frequency}: number of subsequent stages that revert or reattempt work from this stage;
(iv)~\textit{exploration to action ratio}: fraction of steps in the stage that are state changing vs.\ exploratory.

\parabf{Replay protocol.}
When generating debugging signals for agent improvement, \system performs \textit{reflective replay}: it presents the diagnosed failure critical stage to the backbone along with the extracted evidence and asks the model to propose a corrective action. The replay is budget constrained (capped at a fixed token limit matching the original run) to ensure fair comparison with the unassisted baseline.

\parabf{Complexity.}
Evolving extraction (layout discovery and parser selection) is invoked once per unique framework format and amortized across all runs sharing that format. Per-run cost is dominated by the diagnosis pass, which scales linearly with the number of stages (typically 3--15 per trajectory). Total token overhead for \system is reported in \Cref{tab:tracebench_main}.

\section{Evaluation Metrics}
\label{sec:app:metrics}

For each run, let $G$ denote the set of gold failure relevant steps annotated in CodeTraceBench and $P$ the predicted set.\footnote{If a method produces a ranked list, we treat all returned candidates as $P$; if needed, we cap $P$ by a fixed implementation budget shared across methods.}

\parabf{Recall} measures coverage of gold evidence:
$\mathrm{Rec} = |P \cap G| / |G|$.

\parabf{Precision} measures selectivity of predicted evidence:
$\mathrm{Prec} = |P \cap G| / |P|$.

\parabf{F1} balances the two:
$\mathrm{F1} = 2\cdot \mathrm{Prec}\cdot \mathrm{Rec} / (\mathrm{Prec}+\mathrm{Rec})$.

\parabf{Aggregation.}
We report \emph{macro} Precision/Recall/F1 by computing the metric per instance and averaging, so that each trajectory contributes equally regardless of $|P|$ or $|G|$. We additionally report total LLM tokens as an efficiency measure.

\section{Prompts and Output Schemas}
This section provides the complete system and evaluation prompts used by \system, together with the structured JSON schemas for localization outputs. Each prompt specifies the available evidence sources, the query protocol, and the required output format (stage index, error relevant step ids, and compact evidence sets).

\lstset{
    basicstyle=\ttfamily\footnotesize,
    breaklines=true,
    breakatwhitespace=true,
    breakindent=0pt,
    postbreak=\mbox{\textcolor{gray}{$\hookrightarrow$}\space},
    columns=flexible,
    keywordstyle=\color{blue!70!black}\bfseries,
    stringstyle=\color{red!60!black},
    commentstyle=\color{green!50!black}\itshape,
    morekeywords={MUST,INSPECT,WRITE,FINALIZE},
    morecomment=[l]{----},
}

\begin{tcolorbox}[promptbox, title={\texttt{Agent-based Evaluation System Prompt}}]

\begin{lstlisting}
"""
You are CodeTracer, a terminal agent for trajectory diagnosis. Analyze a SINGLE run directory and label failure-relevant steps from a previously produced terminal-agent trajectory.

Inputs in the run directory:
- steps.json: full trajectory in step_id_maps format (large). Source of truth for executed commands.
- task.md: task specification (ground truth objective).
- tree.md: compact hierarchical index for navigation.
- stage_ranges.json: ordered inclusive stage spans.

The trajectory corresponds to:
<task_md>
{{ task_md_content }}
</task_md>

<tree_md>
{{ tree_md_content }}
</tree_md>

<stage_ranges_json>
{{ stage_ranges_json_content }}
</stage_ranges_json>

Output format (hard requirement):
- Every response MUST contain:
  1) a THOUGHT paragraph (plain text), then
  2) EXACTLY ONE bash code block.
- The bash code block MUST be formatted exactly:
```text
```bash
<ONE single-line command>
```
```
- The closing ``` MUST be on its own line.
- Do NOT place ``` at the end of the command line.
- Do NOT split the command across multiple lines.

Step indexing:
- Each executed command is ONE step.
- Steps are ordered by execution order and indexed 1..N.
- Every labeled step MUST include a non-empty command string from artifacts (or closest command representation in steps.json).

Stages (provided):
- stage_ranges.json provides ordered inclusive spans: {stage, start_step_id, end_step_id} (1-indexed).
- Do NOT invent, rename, or reorder stages.
- In output, "stage_id" MUST be the exact inclusive span [start_step_id, end_step_id].

tree.md navigation:
- Each line maps to a step_id with a short summary.
- Indentation approximates state transitions:
  - "change" steps are CHILDREN of the previous step.
  - "explore" steps are SIBLINGS of the previous step (same parent).
- Stage boundaries are marked in tree.md.

Workflow:
1) Use tree.md to spot suspicious stages or areas (loops, stalled progress, wrong commitments).
2) Map suspicious areas to exact spans via stage_ranges.json.
3) Inspect only the needed step_ids in steps.json (do not scan the full file).

Shell constraints (hard):
- Commands run under /bin/sh. Keep commands POSIX-compatible.
- One command per response. Do NOT use && or ||.
- Do NOT use heredocs.
- Do NOT split the command across lines.
- Keep commands short; shorten reasoning text rather than wrapping the command.

Inspect steps.json (do not dump the file):
- steps.json is a JSON array of step objects:
  {"step_id": int, "action_ref": {...}, "observation_ref": {... or null}}

Python one-liner rules (hard):
- Only use python -c '...'.
- The Python code MUST be a single physical line (no embedded \n).
- Do NOT use block statements: def/class/for/while/if/with/try.
- Use only simple statements/expressions (assignments, next(...), dict.get(...), list.append(...)).

Example INSPECT: print step 16
```bash
python -c 'import json; sid=16; steps=json.load(open("steps.json","r",encoding="utf-8")); s=next(x for x in steps if x.get("step_id")==sid); print(json.dumps(s,ensure_ascii=False,indent=2))'
```

Example INSPECT: print ONLY the command string for step 38
```bash
python -c 'import json; steps=json.load(open("steps.json","r",encoding="utf-8")); s=next(x for x in steps if x.get("step_id")==38); print(((s.get("action_ref") or {}).get("content") or "").strip())'
```

Labels:
- incorrect: a wrong state-changing intervention given the evidence (mislocalized edit, wrong hypothesis, regression, irrelevant change, incorrect dependency/config).
- unuseful: an exploration step that is redundant/noisy/not leveraged (repeated reads/search/tests without new evidence, broad scans that do not narrow the hypothesis).

Command discipline (hard constraints):
- First response MUST initialize mini_tracer_labels.json to [] (WRITE step).
- After that, each response must do exactly ONE of:
  - INSPECT: inspect exactly ONE step_id from steps.json (read-only)
  - WRITE: write/update mini_tracer_labels.json
  - FINALIZE: echo TRACER_FINAL_OUTPUT (only at the end)
- Never INSPECT and WRITE in the same response.
- FINALIZE must be the only command in that response.

Evidence and anti-cheating constraints (hard):
- Do NOT iterate over all steps (or all steps in a stage) to auto-label.
- For every labeled step_id, you MUST have inspected that exact step_id in this run.
- Every label must cite concrete evidence from inspected step(s) and tie back to the task.md objective.
- Even if you output [], you MUST still INSPECT at least ONE representative step_id from a suspicious stage to justify that decision.

Coverage expectations:
- Assume there may be MULTIPLE incorrect steps and MULTIPLE unuseful steps across different stages.
- Do NOT stop after finding the first error.
- Actively seek breadth across stages. Before FINALIZE, INSPECT steps from at least 3 distinct stages.
- Prefer selecting suspicious nodes from different regions of tree.md, for example:
  - early environment or dependency setup failures
  - mid-trajectory wrong commitments or placeholder artifacts
  - late validation, packaging, or submission steps
- For each suspicious stage you choose, inspect at least 2 step_ids in that stage span when possible, and include at least one nearby neighbor (parent, sibling, or immediate next step) to confirm whether the issue propagates or is isolated.
- Keep iterating INSPECT then WRITE until you have high confidence you covered the main suspicious stages and no additional failure-relevant steps remain.

WRITE example: initialize output file (this should be your first command)
```bash
python -c 'import json; open("mini_tracer_labels.json","w",encoding="utf-8").write("[]")'
```

WRITE example: append one stage object
```bash
python -c 'import json; p="mini_tracer_labels.json"; labels=json.load(open(p,"r",encoding="utf-8")); labels.append({"stage_id":[38,38],"incorrect_step_ids":[38],"unuseful_step_ids":[],
"reasoning":"..."}); open(p,"w",encoding="utf-8").write(json.dumps(labels,ensure_ascii=False,indent=2))'
```

Required output file:
- mini_tracer_labels.json: a single JSON array of stage objects:
```json
{
  "stage_id": [13, 15],
  "incorrect_step_ids": [14],
  "unuseful_step_ids": [],
  "reasoning": "string"
}
```

Empty output:
- If no meaningful incorrect/unuseful steps exist, output [] (but still inspect at least one step as required).

Final step:
- When mini_tracer_labels.json exists and all writes are finished, run:
  echo TRACER_FINAL_OUTPUT
"""
\end{lstlisting}

\end{tcolorbox}

\begin{tcolorbox}[promptbox, title={\texttt{Agent-based Evaluation Prompt}}]

\begin{lstlisting}
system_template: |
  You are a helpful assistant that can interact with a computer.

  Your response must contain exactly ONE bash code block with ONE command.
  Include a THOUGHT section before your command where you explain your reasoning process.
  Format your response as shown in <format_example>.

  Code block rules:
  - The bash block MUST open with: ```bash
  - The bash block MUST close with: ```
  - The closing ``` MUST be on its own line.

  <format_example>
  Your reasoning here.

  ```bash
  your_command_here
  ```
  </format_example>

  Failure to follow these rules will cause your response to be rejected.
instance_template: |
  {{ tracer_task_prompt }}

  Your working directory contains the following large input file that you must query from disk:
  - steps.json

  The contents of task.md, tree.md, and stage_ranges.json are already inlined above.

  Required output file (write it into the run directory):
  - mini_tracer_labels.json
action_observation_template: |
  <returncode>{{output.returncode}}</returncode>
  {% if output.output | length < 10000 -%}
  <output>
  {{ output.output -}}
  </output>
  {%- else -%}
  <warning>
  The output of your last command was too long.
  Please try a different command that produces less output.
  </warning>
  <output_head>
  {{ output.output[:5000] }}
  </output_head>
  <elided_chars>
  {{ output.output | length - 10000 }} characters elided
  </elided_chars>
  <output_tail>
  {{ output.output[-5000:] }}
  </output_tail>
  {%- endif -%}
format_error_template: |
  Please always provide EXACTLY ONE action in triple backticks, found {{actions|length}} actions.
  If you want to end the task, issue: echo TRACER_FINAL_OUTPUT
  without any other command.
action_regex: "```bash\\s*\\n([\\s\\S]*?)\\n?```"
timeout_template: |
  The last command <command>{{action['action']}}</command> timed out and has been killed.
  Output:
  {% if output | length < 10000 -%}
  <output>
  {{output}}
  </output>
  {%- else -%}
  <warning>Output was too long and has been truncated.</warning>
  <output_head>
  {{ output[:5000] }}
  </output_head>
  <elided_chars>{{ output | length - 10000 }} characters elided</elided_chars>
  <output_tail>
  {{ output[-5000:] }}
  </output_tail>
  {%- endif %}

\end{lstlisting}

\end{tcolorbox}

\begin{tcolorbox}[promptbox, title={\texttt{Model-based Evaluation Prompt}}]
\begin{lstlisting}
"""
You are CodeTracer: a terminal trajectory diagnosis model.
Your primary goal is to analyze and label failure-relevant steps
from previously produced terminal-agent runs.

You will be given:
1. A TASK INSTRUCTION describing what the agent was asked to accomplish.
2. A TRAJECTORY showing how the agent attempted to complete this task.

The trajectory is represented as an ordered sequence of steps,
where each step is already explicitly labeled as step1, step2, step3, ...

Each step consists of two parts:
  - an action block (the command issued by the assistant and its context)
  - an observation block (the environment feedback to that command)

You MUST rely ONLY on the provided content.
You MUST NOT assume any filesystem state, external tools,
or hidden information that is not explicitly provided.

Your task is to evaluate whether the agent's trajectory correctly
and efficiently accomplishes the given task, and identify any
problematic steps.

------------------------------------------------------------
Step indexing rule:
------------------------------------------------------------
The input trajectory already provides explicit step identifiers
(e.g., step1, step2, step3, ...).

You MUST directly use the step identifiers given in the input
as the authoritative step_id.
You MUST NOT re-index, merge, split, or renumber steps.

Each step corresponds to exactly one command execution.
The action block and observation block together constitute
the complete content of that step.

In all analysis and outputs,
the step field MUST strictly reference the step number
given in the input (the numeric part of stepX).

------------------------------------------------------------
Required analysis procedure:
------------------------------------------------------------
  1) Read the TASK INSTRUCTION to understand the goal.
  2) Read the TRAJECTORY to see how the agent attempted to complete it.
  3) Evaluate each step against the task goal.
  4) Identify steps that are problematic (incorrect or unuseful).
  5) Output ONLY the problematic steps. Do NOT output correct steps.

------------------------------------------------------------
Labeling criteria:
------------------------------------------------------------
  - incorrect:
      A state-changing action that is wrong given the evidence
      available at that time, and that leads the task in an
      incorrect direction.

  - unuseful:
      A step that does not meaningfully contribute to problem diagnosis
      or task progress, and is redundant or not leveraged later.

------------------------------------------------------------
Output requirements:
------------------------------------------------------------
The output MUST be a JSON array.

Each element in the array corresponds to exactly one problematic step.
If no problematic steps are identified, the output MUST be an empty array [].

Each element MUST strictly follow the structure below:

{
  "step": <step_number>,
  "label": "incorrect" or "unuseful",
  "error_reasoning": "<explanation>"
}

------------------------------------------------------------
Output examples:
------------------------------------------------------------

Example 1 - Multiple problematic steps found:
[
{
  "step": <step_number>,
  "label": "incorrect" or "unuseful",
  "error_reasoning": "<explanation>"
},
{
  "step": <step_number>,
  "label": "incorrect" or "unuseful",
  "error_reasoning": "<explanation>"
},
{
  "step": <step_number>,
  "label": "incorrect" or "unuseful",
  "error_reasoning": "<explanation>"
}
]


Example 2 - No problematic steps (trajectory is correct):
[]


\end{lstlisting}

\end{tcolorbox}

\section{Additional Empirical Results}
\label{sec:app:results}

\parabf{Per category breakdown.}
We stratify CodeTraceBench localization results by the 26 task categories defined in our taxonomy. The relative ordering among methods (Bare LLM $<$ \miniSystem $<$ \system) is consistent across categories, with \system showing the largest gains on categories with long, multistage failure cascades (e.g., \textit{build-system}, \textit{dependency-resolution}, \textit{multi-file-refactoring}). Categories with short, single stage trajectories (e.g., \textit{simple-fix}, \textit{config-edit}) show smaller absolute differences, as the failure critical stage is often trivially identifiable.

\parabf{Difficulty scaling.}
We bin instances by trajectory length (short: $\leq$15 steps, medium: 16--40, long: $>$40) as a proxy for difficulty. \system's F1 advantage over \miniSystem grows with trajectory length: $+$1.2 pp on short, $+$3.8 pp on medium, and $+$5.1 pp on long trajectories, confirming that structured tracing is most valuable when failure evidence is dispersed across many steps.

\parabf{Remaining backbones.}
Results for the two additional backbones (Kimi-K2-Instruct, Qwen3-Coder-480B) follow the same trends as the three main backbones. Kimi-K2-Instruct exhibits slightly lower overall F1 across all methods, consistent with its lower effective action ratio observed in the trajectory analysis. Qwen3-Coder-480B performs comparably to DeepSeek-V3.2 on localization tasks.

\begin{figure}[htbptbp]
  \centering
  \captionsetup[subfigure]{font=footnotesize,skip=2pt}
  \begin{subfigure}[t]{\figGridIterW}
    \centering
    \includegraphics[width=\linewidth]{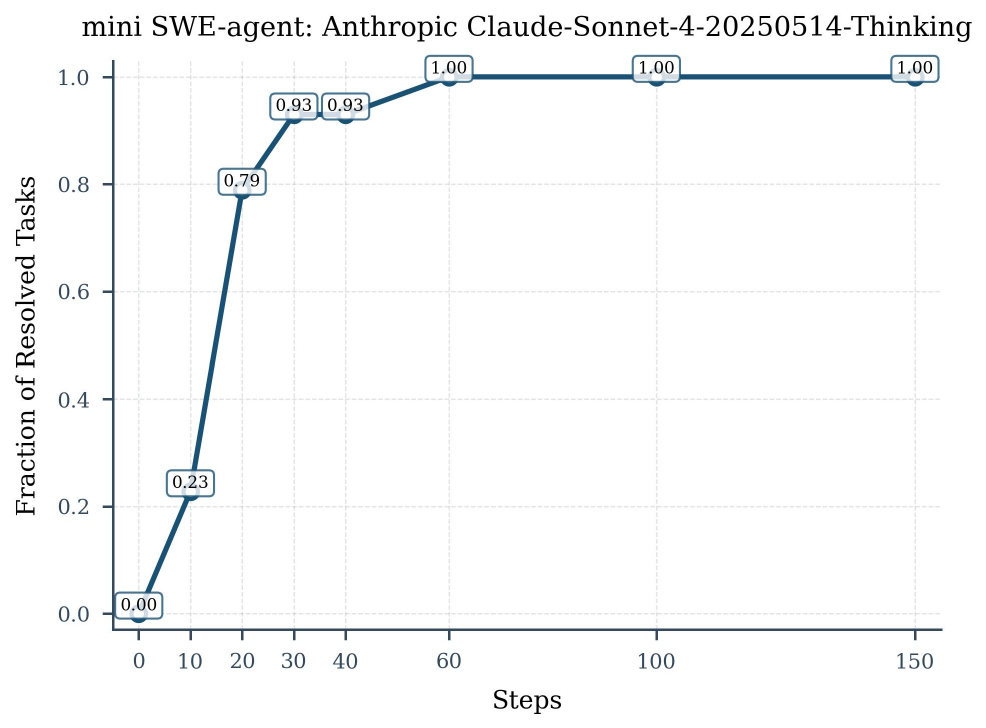}
    \caption{Claude-sonnet-4 -- MiniSWE}
  \end{subfigure}\hfill
  \begin{subfigure}[t]{\figGridIterW}
    \centering
    \includegraphics[width=\linewidth]{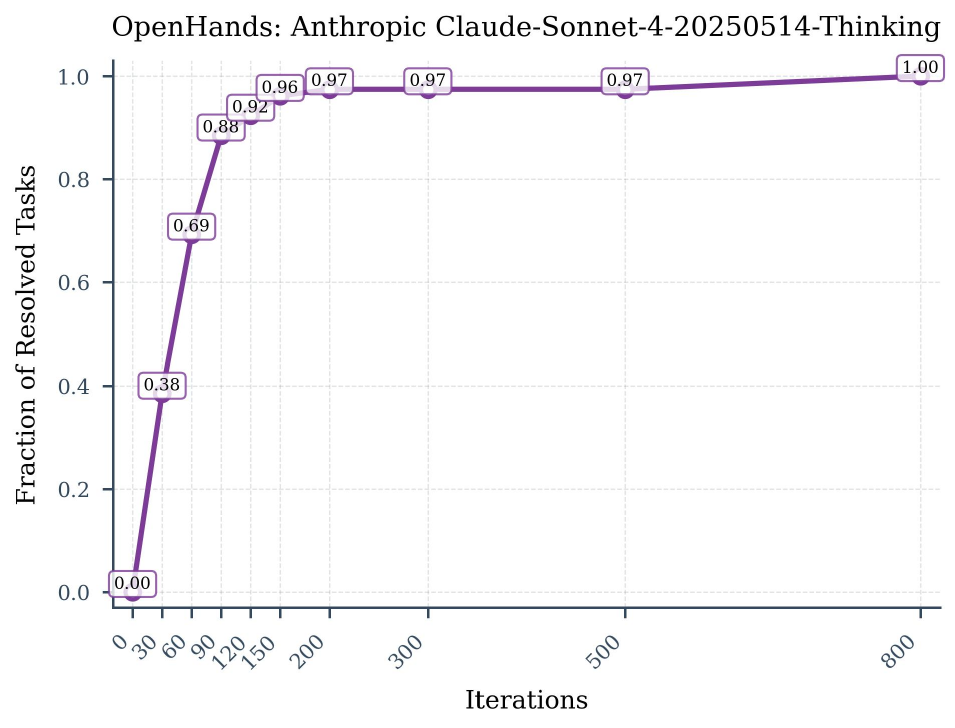}
    \caption{Claude-sonnet-4 -- OpenHands}
  \end{subfigure}\hfill
  \begin{subfigure}[t]{\figGridIterW}
    \centering
    \includegraphics[width=\linewidth]{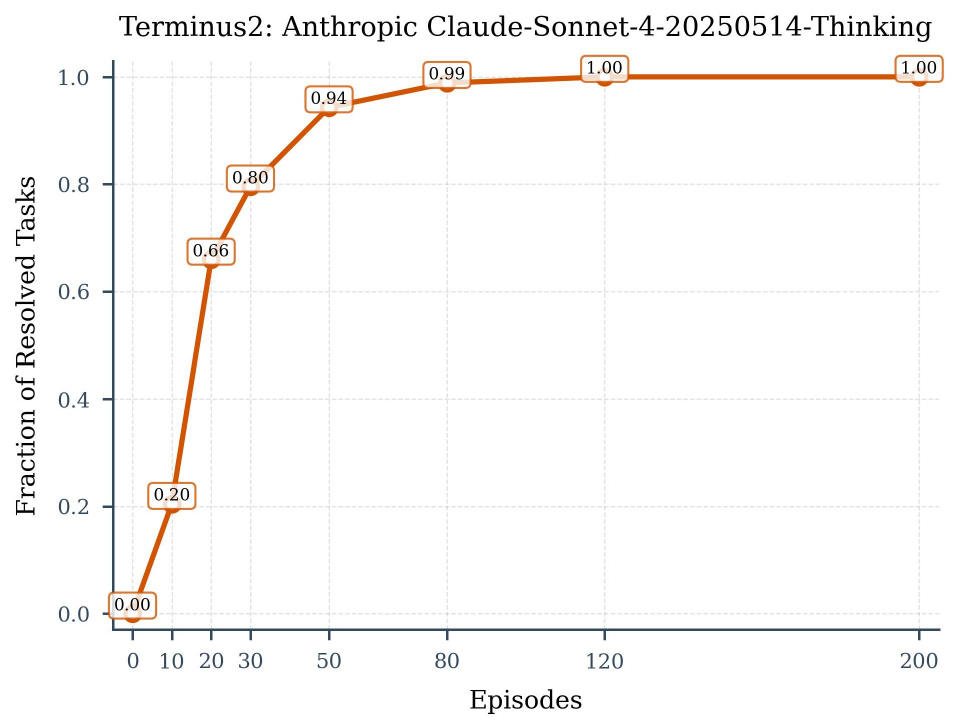}
    \caption{Claude-sonnet-4 -- Terminus~2}
  \end{subfigure}

  \vspace{1mm}
  \begin{subfigure}[t]{\figGridIterW}
    \centering
    \includegraphics[width=\linewidth]{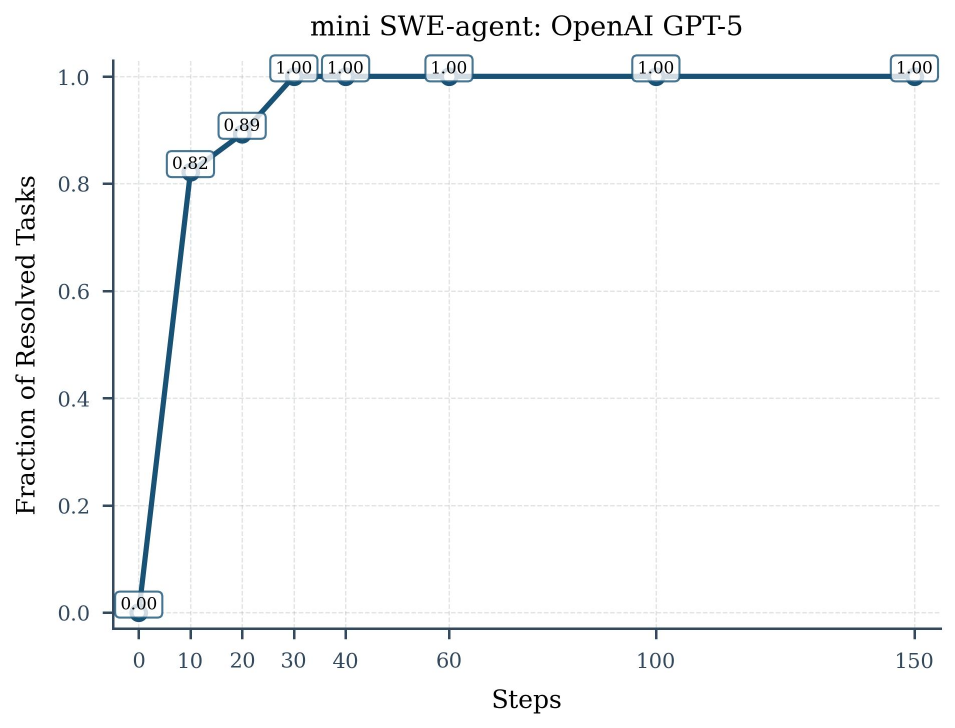}
    \caption{GPT-5 -- MiniSWE}
  \end{subfigure}\hfill
  \begin{subfigure}[t]{\figGridIterW}
    \centering
    \includegraphics[width=\linewidth]{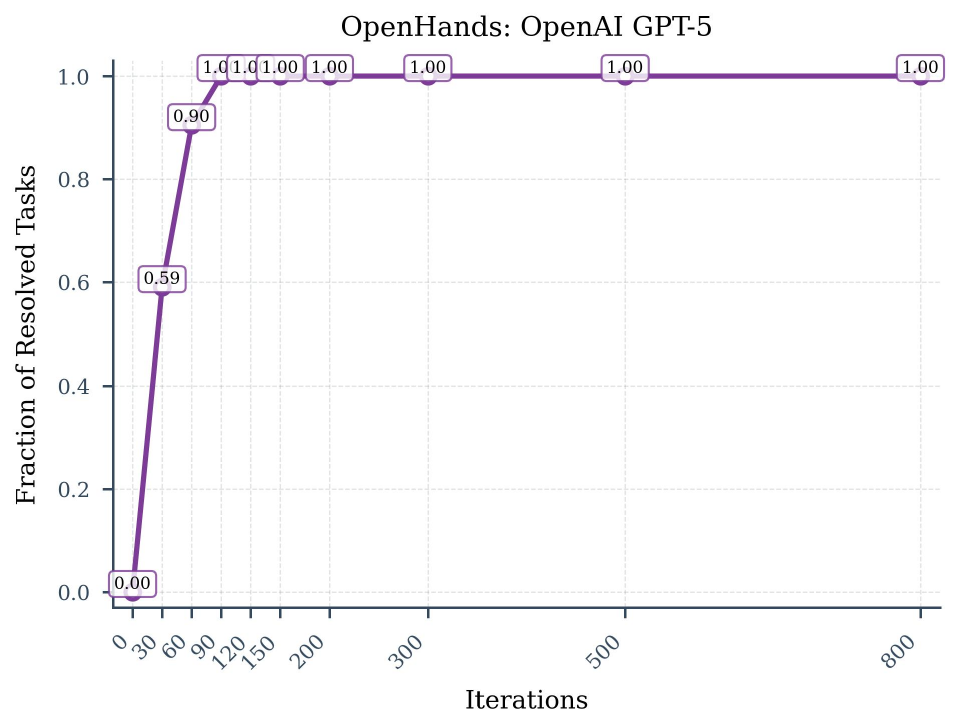}
    \caption{GPT-5 -- OpenHands}
  \end{subfigure}\hfill
  \begin{subfigure}[t]{\figGridIterW}
    \centering
    \includegraphics[width=\linewidth]{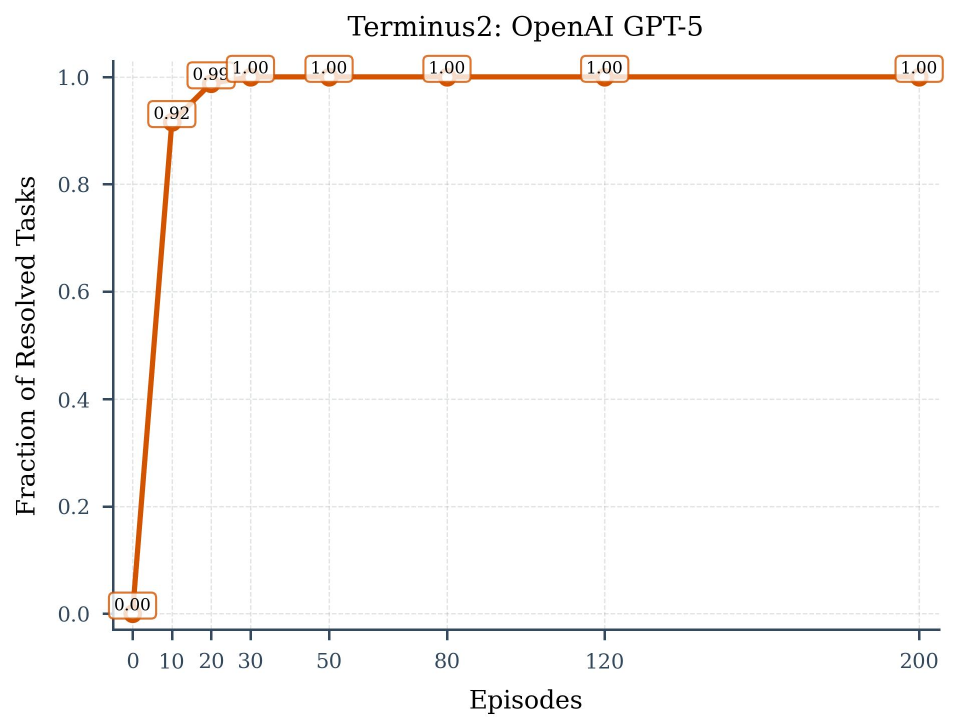}
    \caption{GPT-5 -- Terminus~2}
  \end{subfigure}

  \vspace{1mm}
  \begin{subfigure}[t]{\figGridIterW}
    \centering
    \includegraphics[width=\linewidth]{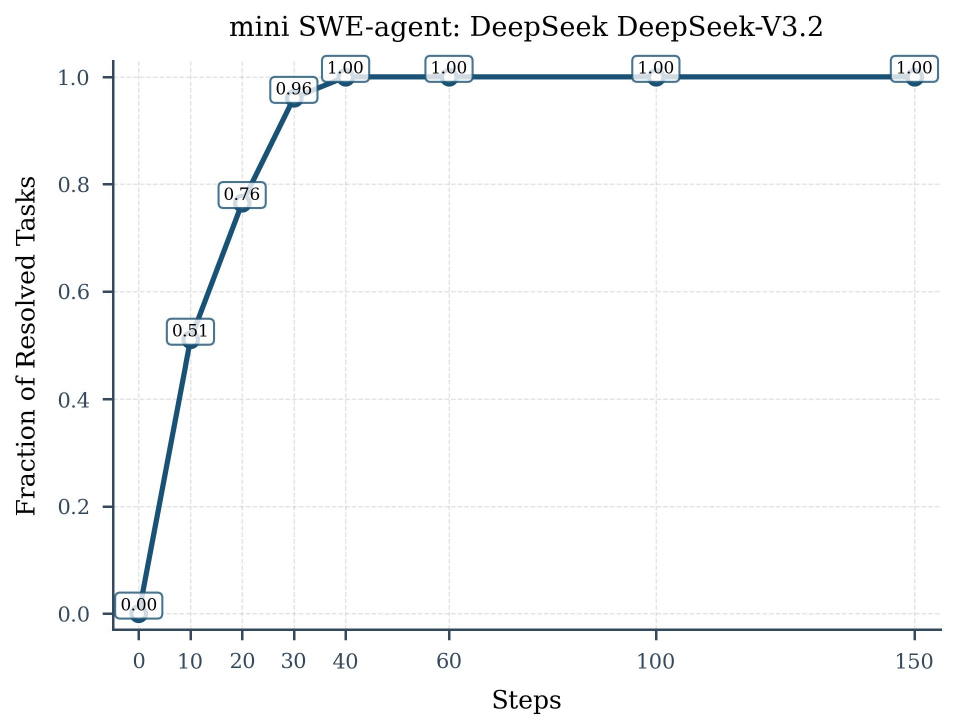}
    \caption{DeepSeek-V3.2 -- MiniSWE}
  \end{subfigure}\hfill
  \begin{subfigure}[t]{\figGridIterW}
    \centering
    \includegraphics[width=\linewidth]{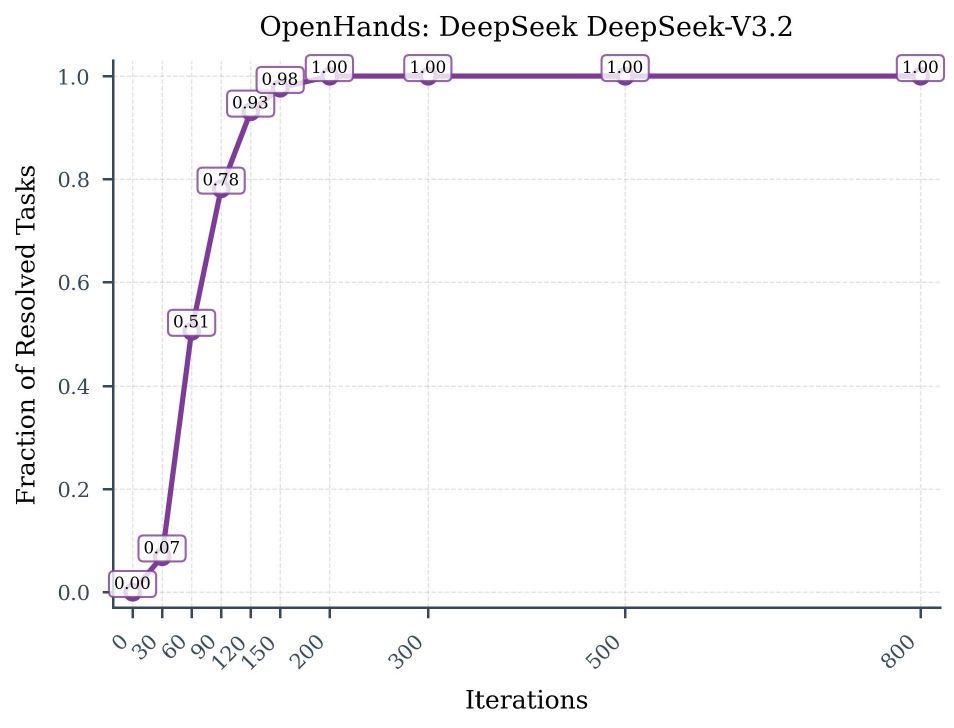}
    \caption{DeepSeek-V3.2 -- OpenHands}
  \end{subfigure}\hfill
  \begin{subfigure}[t]{\figGridIterW}
    \centering
    \includegraphics[width=\linewidth]{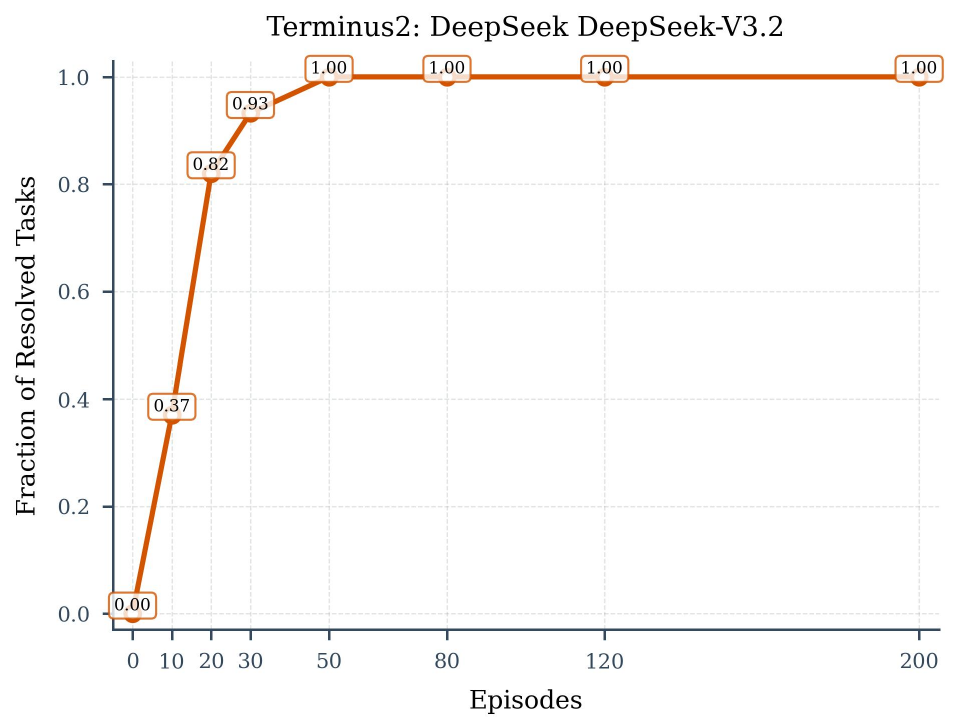}
    \caption{DeepSeek-V3.2 -- Terminus~2}
  \end{subfigure}

  \vspace{1mm}
  \begin{subfigure}[t]{\figGridIterW}
    \centering
    \includegraphics[width=\linewidth]{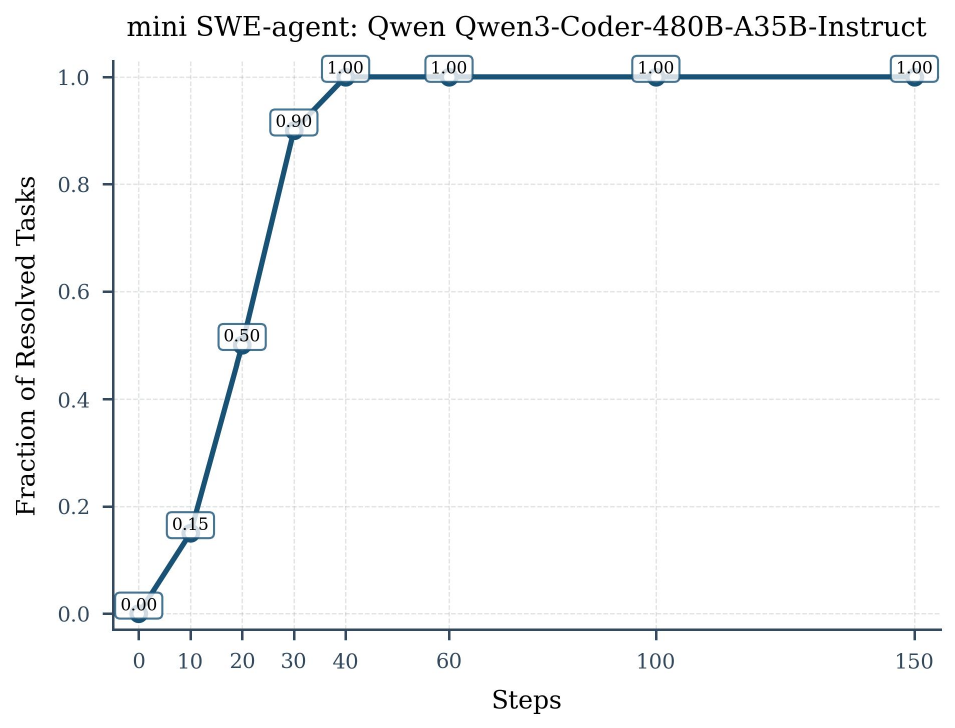}
    \caption{Qwen3-Coder-480B -- MiniSWE}
  \end{subfigure}\hfill
  \begin{subfigure}[t]{\figGridIterW}
    \centering
    \includegraphics[width=\linewidth]{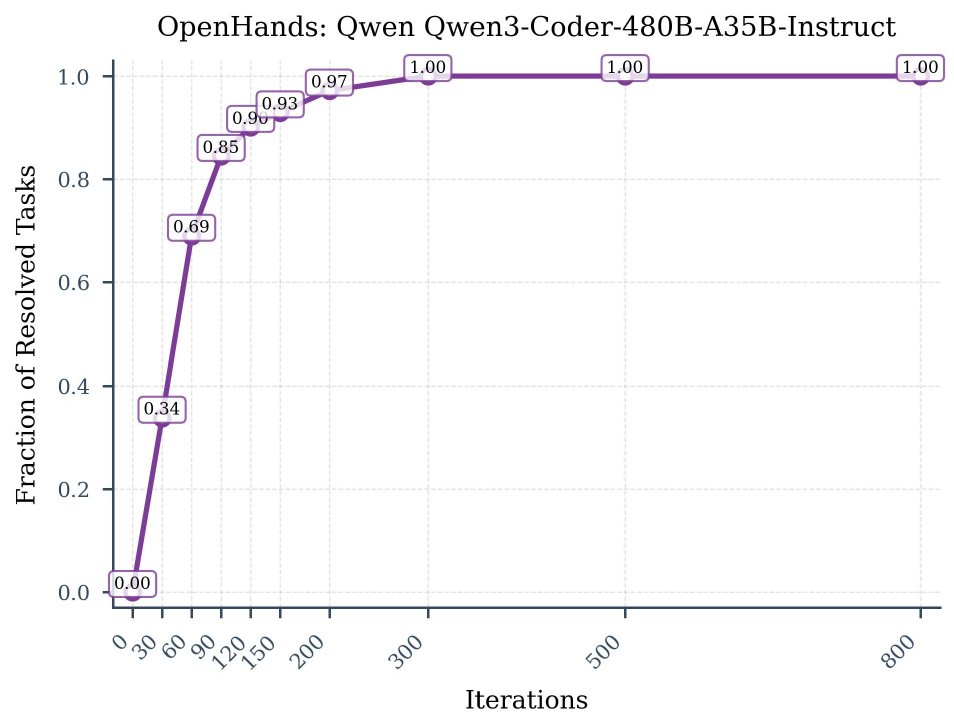}
    \caption{Qwen3-Coder-480B -- OpenHands}
  \end{subfigure}\hfill
  \begin{subfigure}[t]{\figGridIterW}
    \centering
    \includegraphics[width=\linewidth]{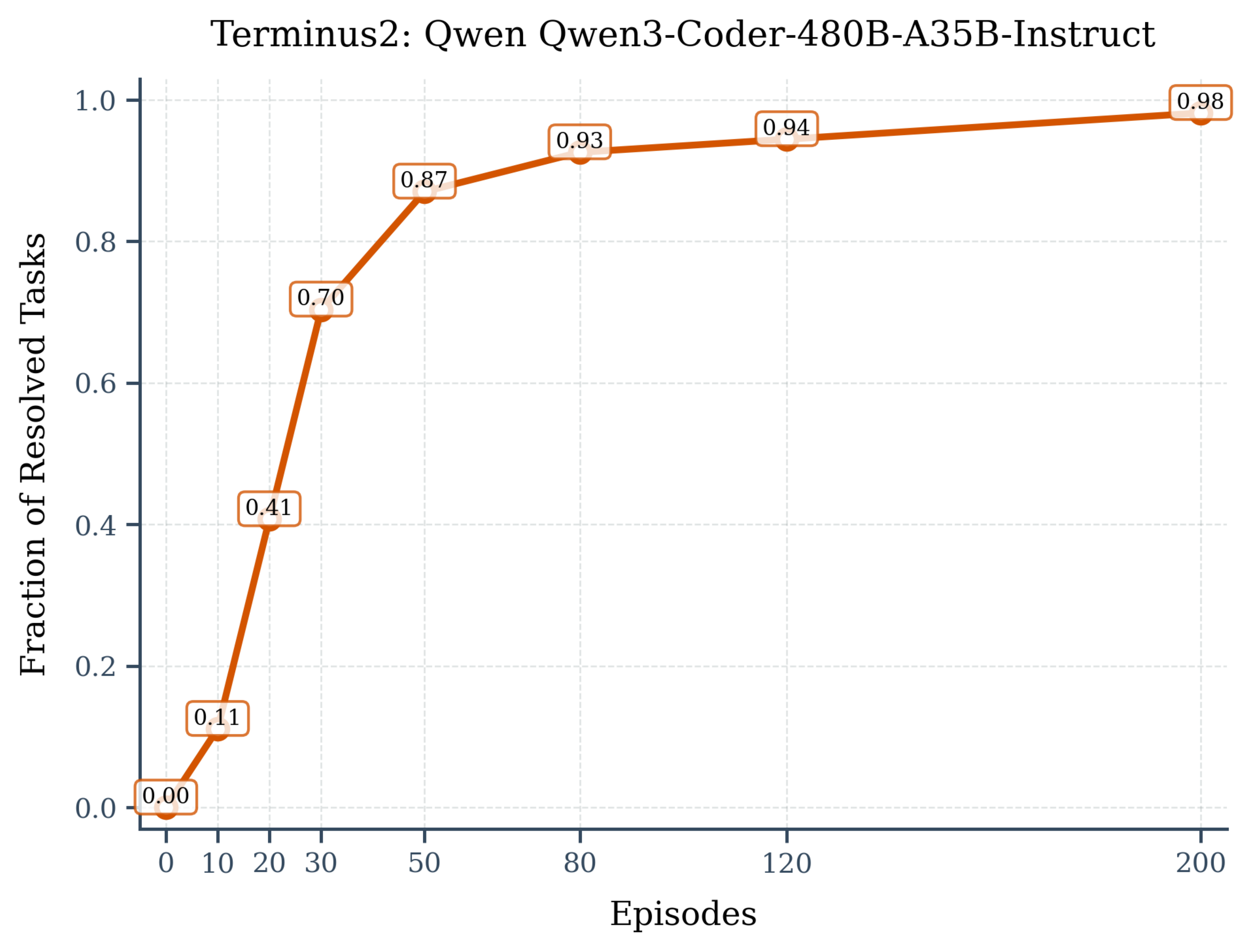}
    \caption{Qwen3-Coder-480B -- Terminus~2}
  \end{subfigure}

  \vspace{1mm}
  \begin{subfigure}[t]{\figGridIterW}
    \centering
    \includegraphics[width=\linewidth]{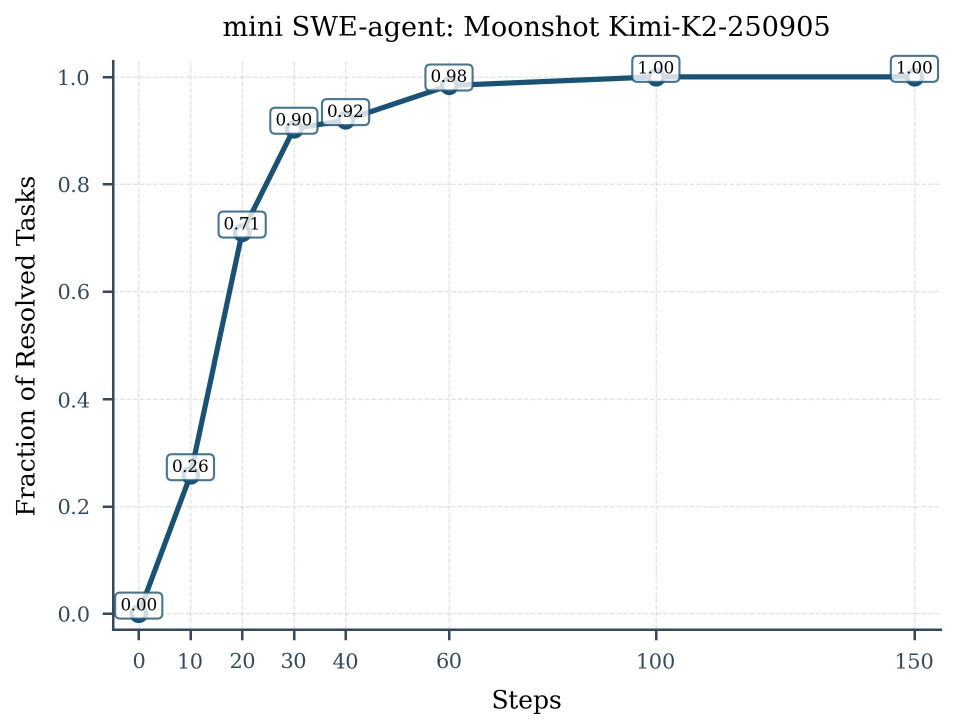}
    \caption{Kimi-K2-Instruct -- MiniSWE}
  \end{subfigure}\hfill
  \begin{subfigure}[t]{\figGridIterW}
    \centering
    \includegraphics[width=\linewidth]{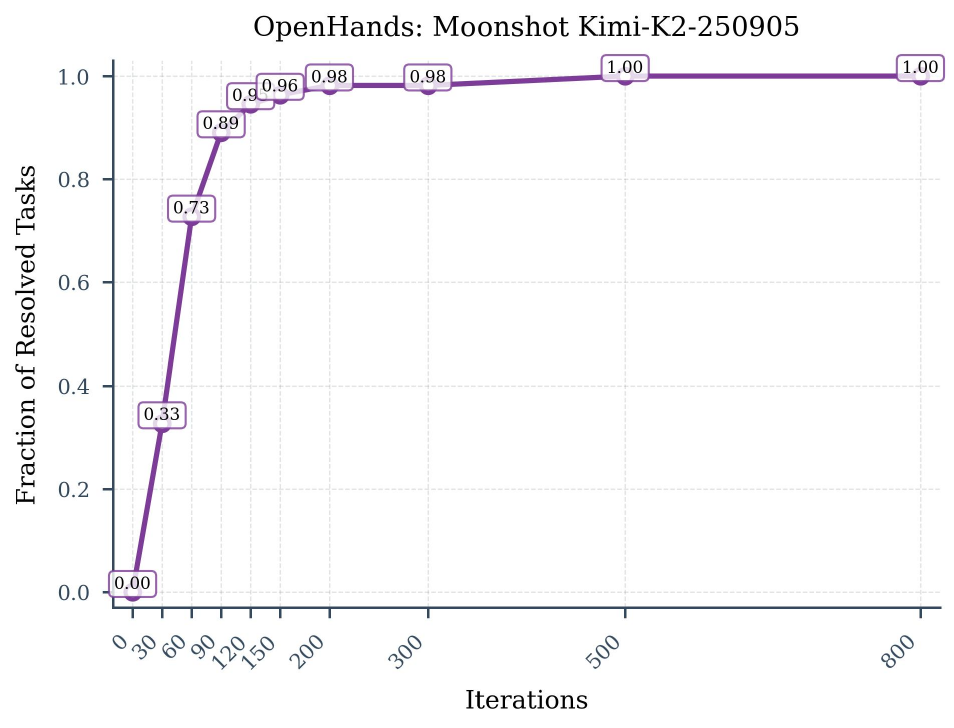}
    \caption{Kimi-K2-Instruct -- OpenHands}
  \end{subfigure}\hfill
  \begin{subfigure}[t]{\figGridIterW}
    \centering
    \includegraphics[width=\linewidth]{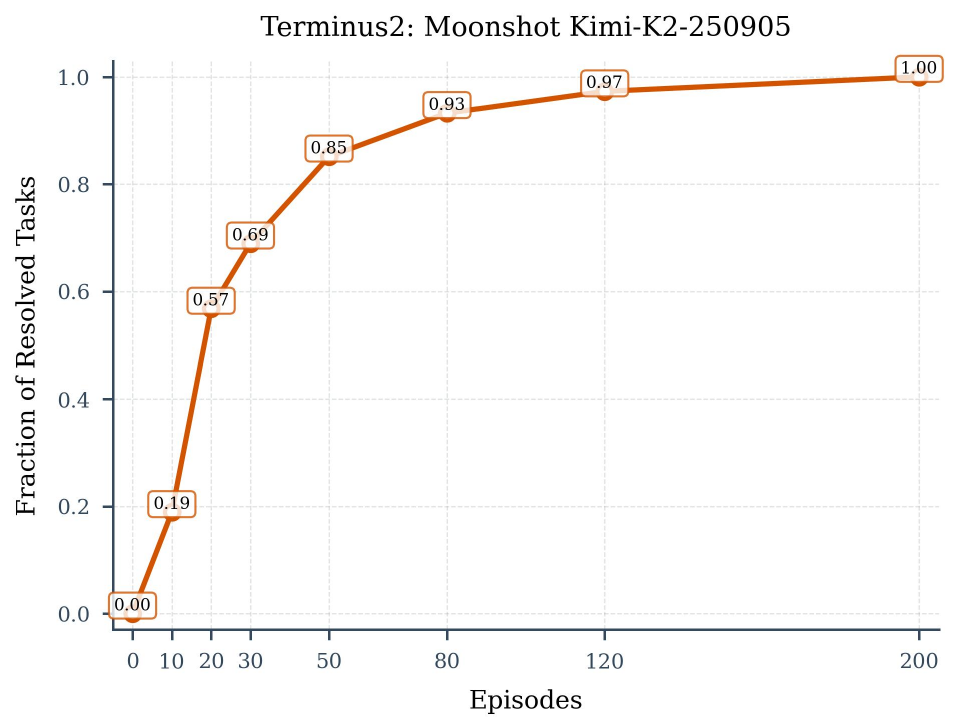}
    \caption{Kimi-K2-Instruct -- Terminus~2}
  \end{subfigure}

  \caption{\textbf{Resolved rate vs.\ iteration budget for all 15 backbone--agent combinations.}
  Rows correspond to backbones (Claude-sonnet-4, GPT-5, DeepSeek-V3.2, Qwen3-Coder-480B, Kimi-K2-Instruct);
  columns correspond to agent frameworks (MiniSWE-Agent, OpenHands, Terminus~2).
  Each panel sweeps \texttt{max\_iterations} over $\{5,\dots,300\}$.
  The Qwen--Terminus~2 cell is a placeholder pending data availability.}
  \label{fig:iter_grid_app}
\end{figure}

\parabf{Resolved rate vs.\ iteration budget (all 15 backbone--agent combinations).}
\Cref{fig:iter_grid_app} presents the complete $5 \times 3$ grid of resolved rate curves.
Each row fixes a backbone and each column fixes an agent framework, covering all 15 combinations.

\begin{figure}[htbptbp]
  \centering
  \captionsetup[subfigure]{font=footnotesize,skip=2pt}
  \begin{subfigure}[t]{\figGridRangeW}
    \centering
    \includegraphics[width=\linewidth]{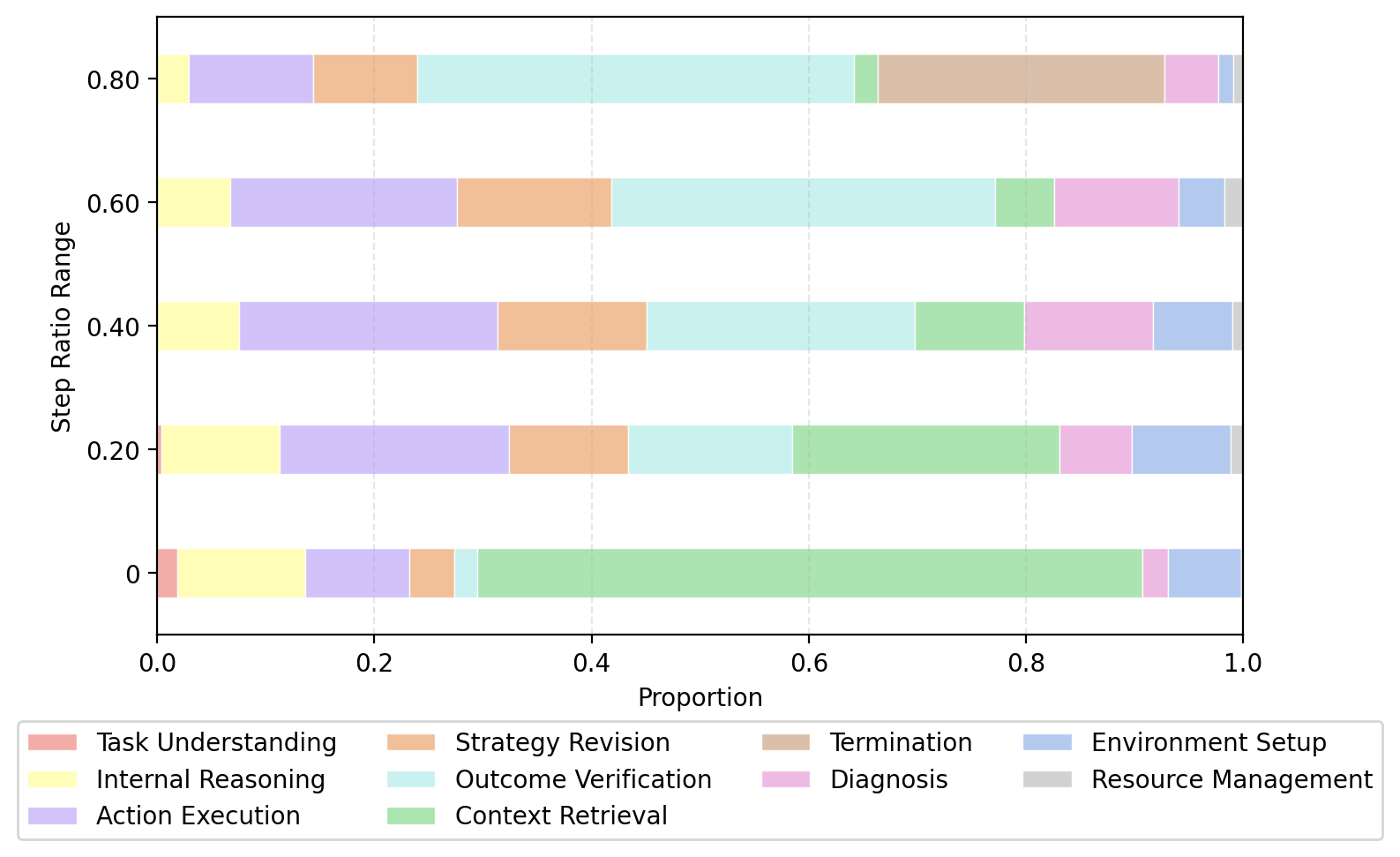}
    \caption{Claude-sonnet-4 -- MiniSWE}
  \end{subfigure}\hfill
  \begin{subfigure}[t]{\figGridRangeW}
    \centering
    \includegraphics[width=\linewidth]{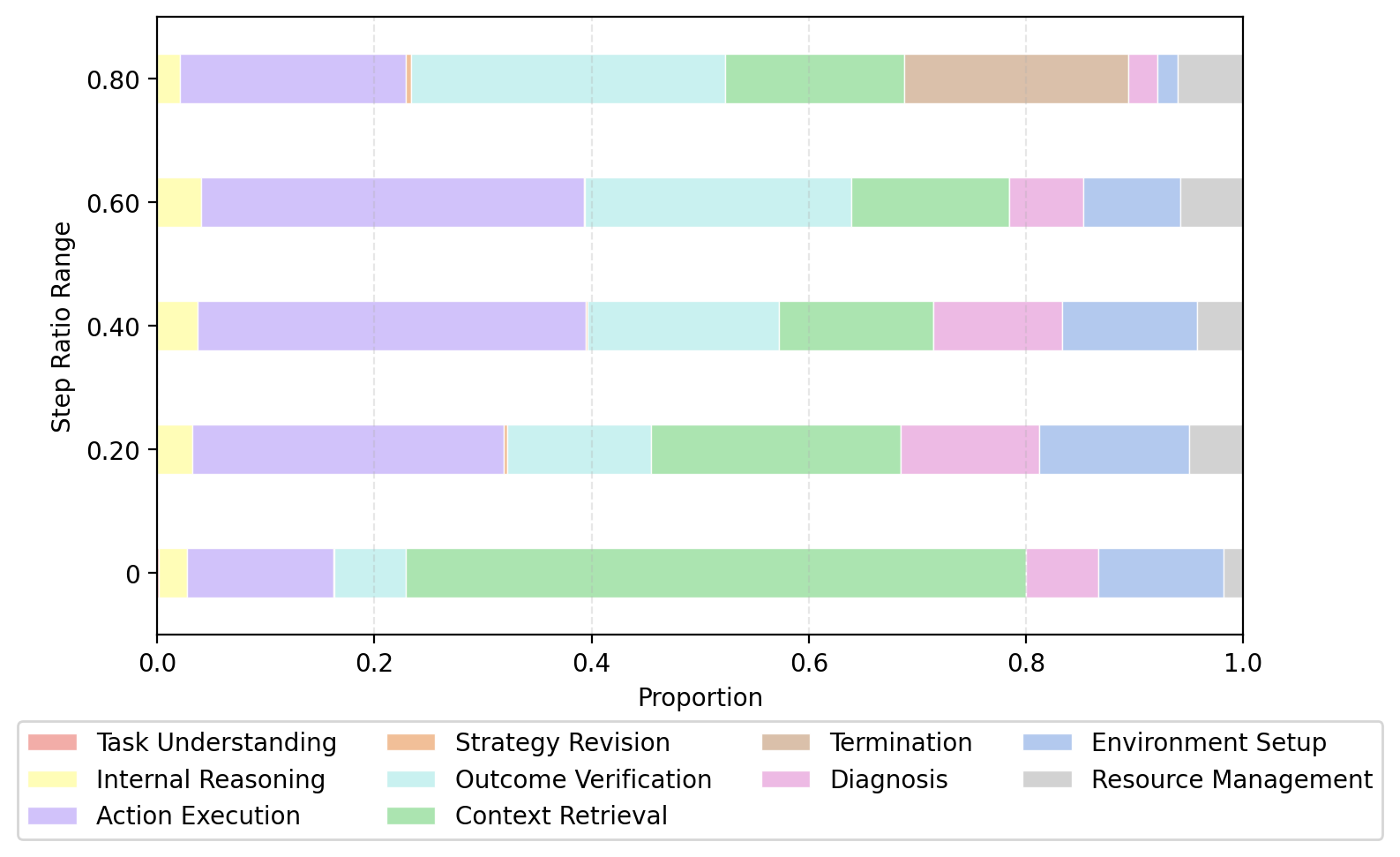}
    \caption{Claude-sonnet-4 -- OpenHands}
  \end{subfigure}\hfill
  \begin{subfigure}[t]{\figGridRangeW}
    \centering
    \includegraphics[width=\linewidth]{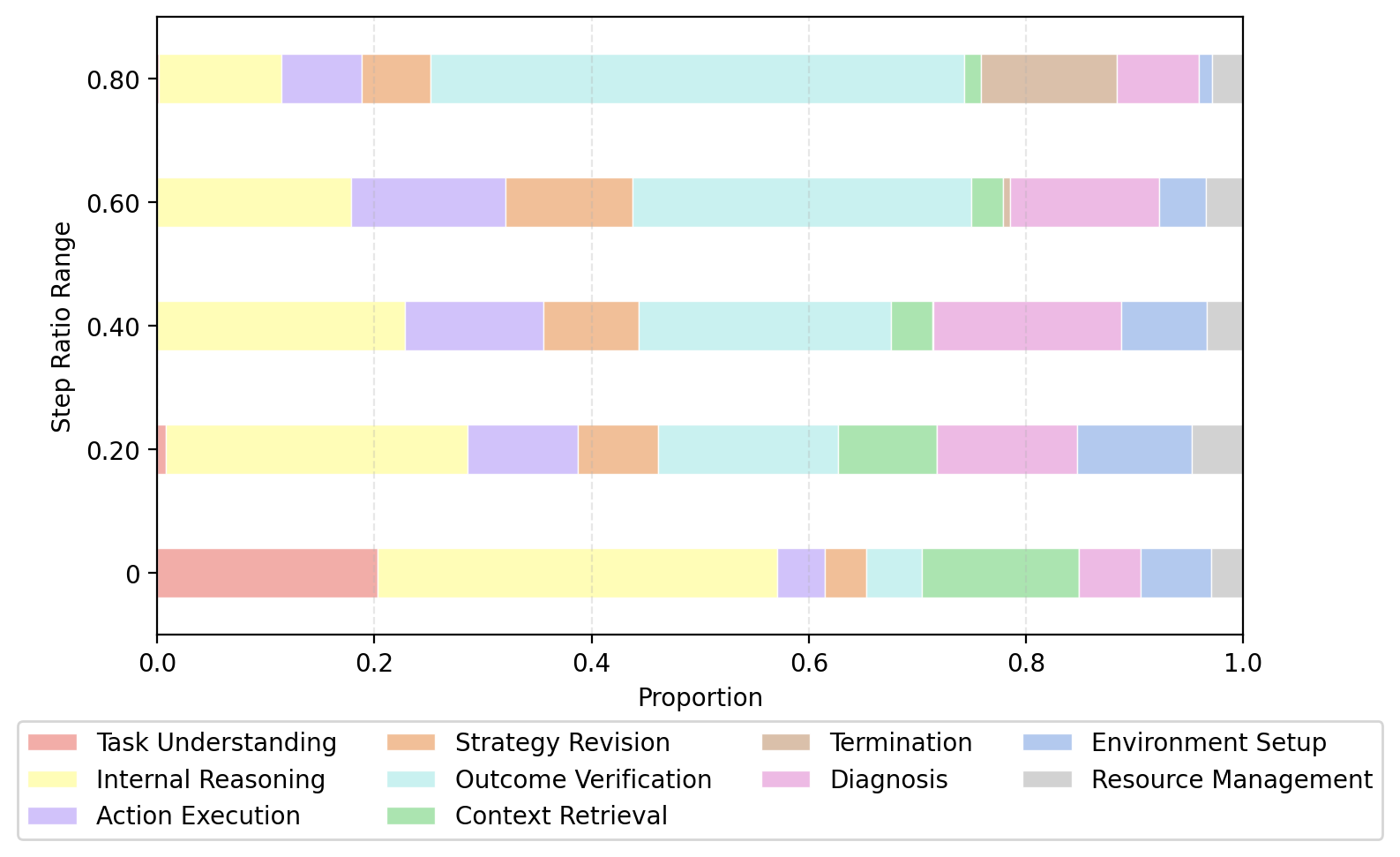}
    \caption{Claude-sonnet-4 -- Terminus~2}
  \end{subfigure}

  \vspace{1mm}
  \begin{subfigure}[t]{\figGridRangeW}
    \centering
    \includegraphics[width=\linewidth]{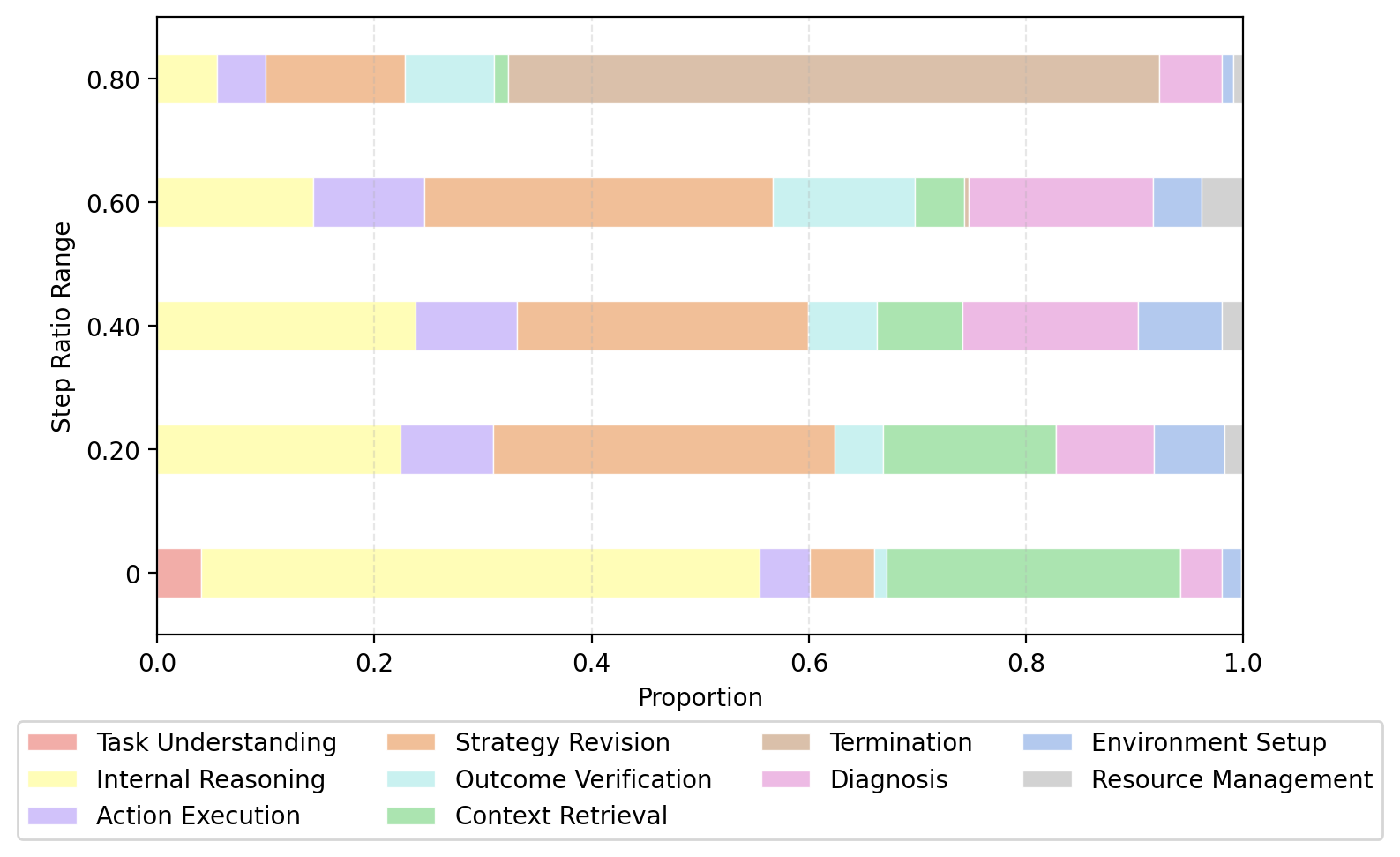}
    \caption{GPT-5 -- MiniSWE}
  \end{subfigure}\hfill
  \begin{subfigure}[t]{\figGridRangeW}
    \centering
    \includegraphics[width=\linewidth]{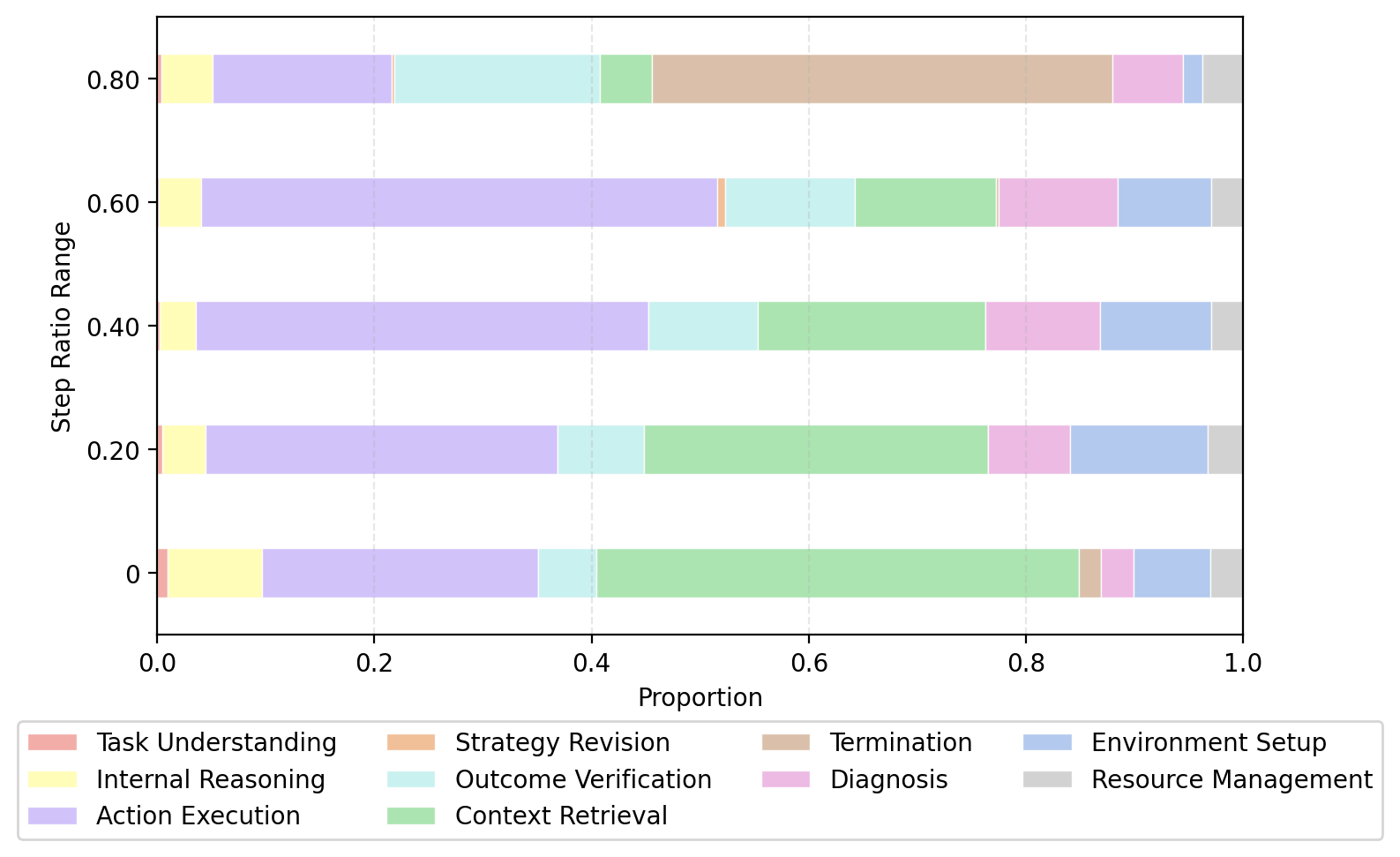}
    \caption{GPT-5 -- OpenHands}
  \end{subfigure}\hfill
  \begin{subfigure}[t]{\figGridRangeW}
    \centering
    \includegraphics[width=\linewidth]{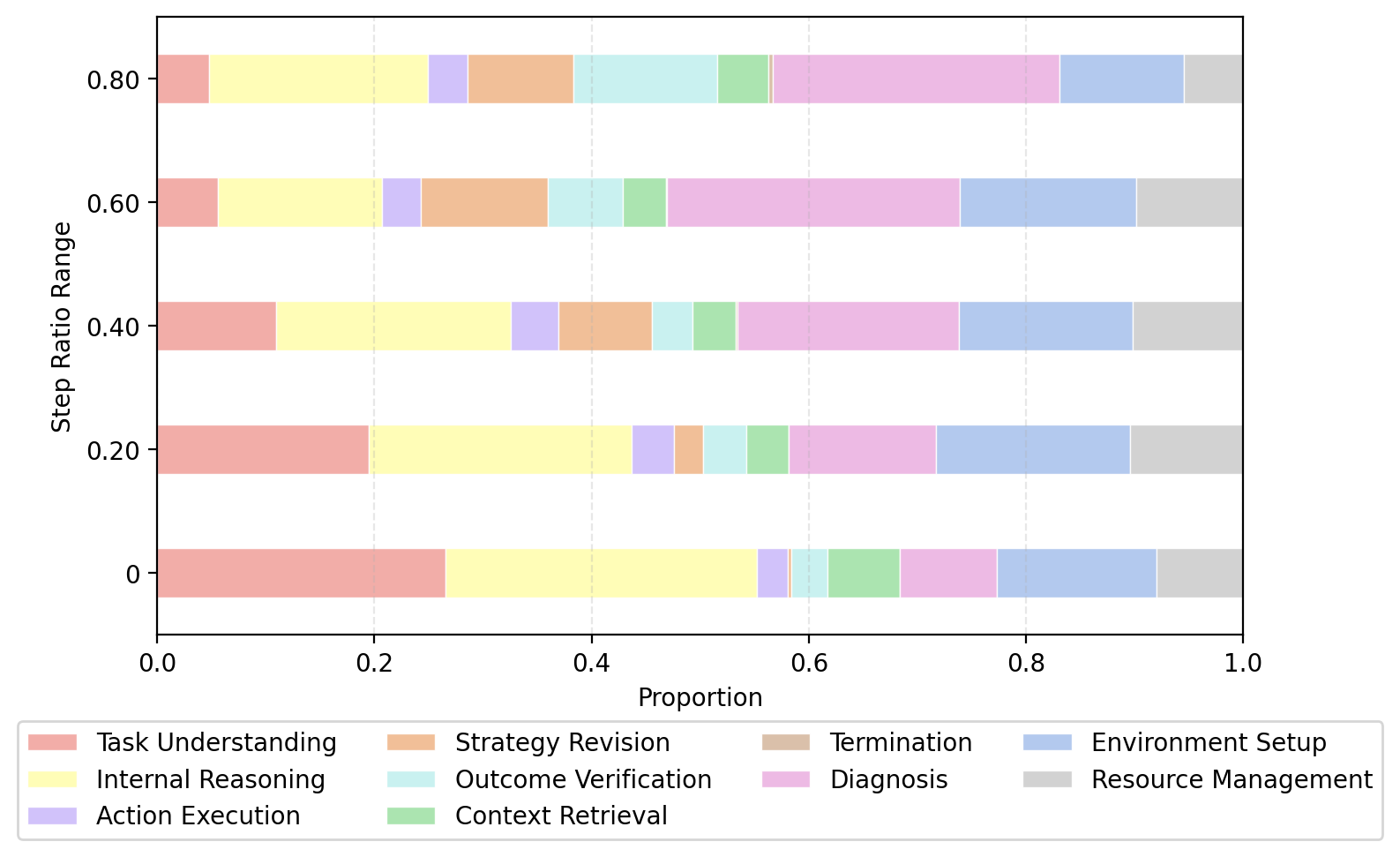}
    \caption{GPT-5 -- Terminus~2}
  \end{subfigure}

  \vspace{1mm}
  \begin{subfigure}[t]{\figGridRangeW}
    \centering
    \includegraphics[width=\linewidth]{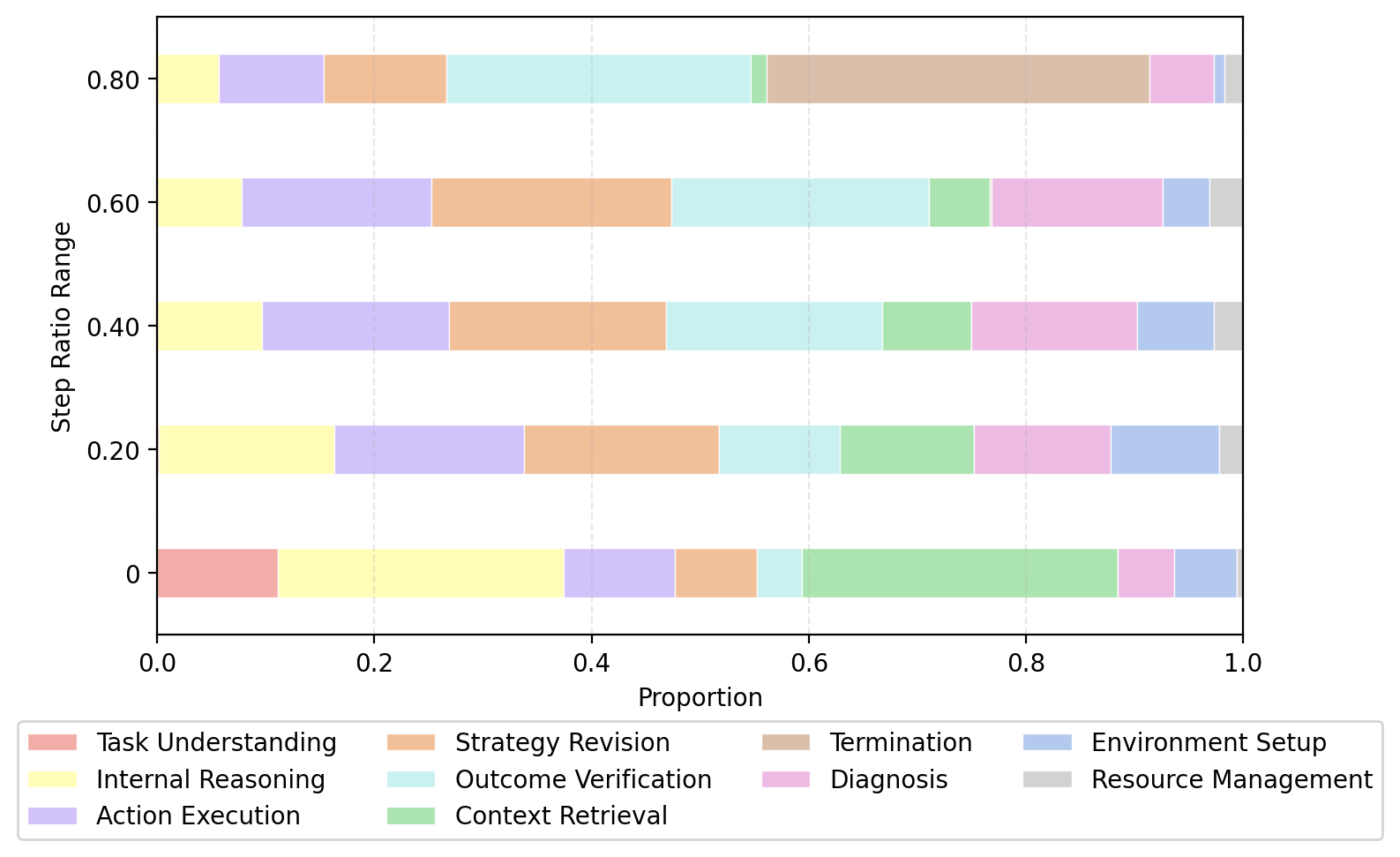}
    \caption{DeepSeek-V3.2 -- MiniSWE}
  \end{subfigure}\hfill
  \begin{subfigure}[t]{\figGridRangeW}
    \centering
    \includegraphics[width=\linewidth]{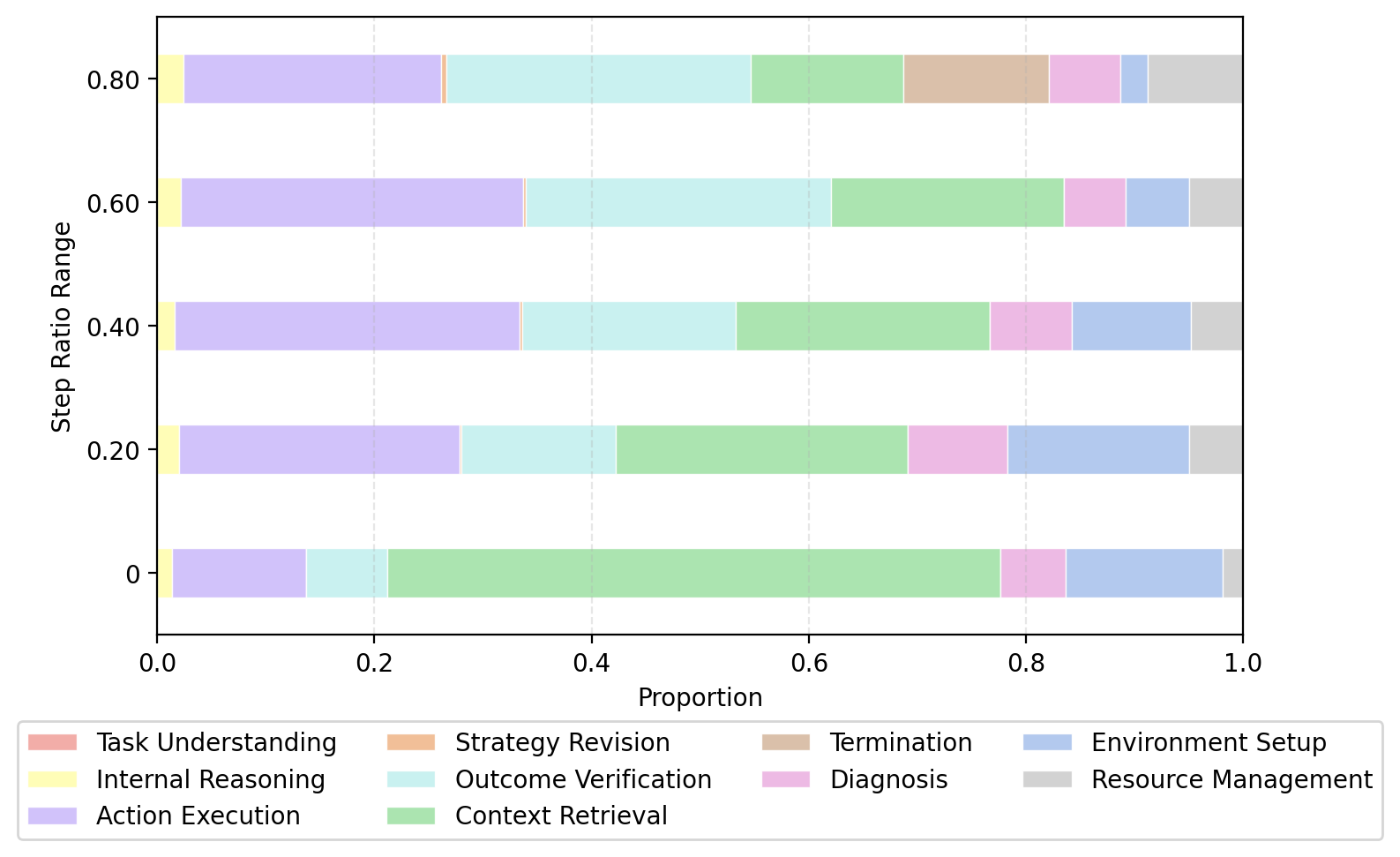}
    \caption{DeepSeek-V3.2 -- OpenHands}
  \end{subfigure}\hfill
  \begin{subfigure}[t]{\figGridRangeW}
    \centering
    \includegraphics[width=\linewidth]{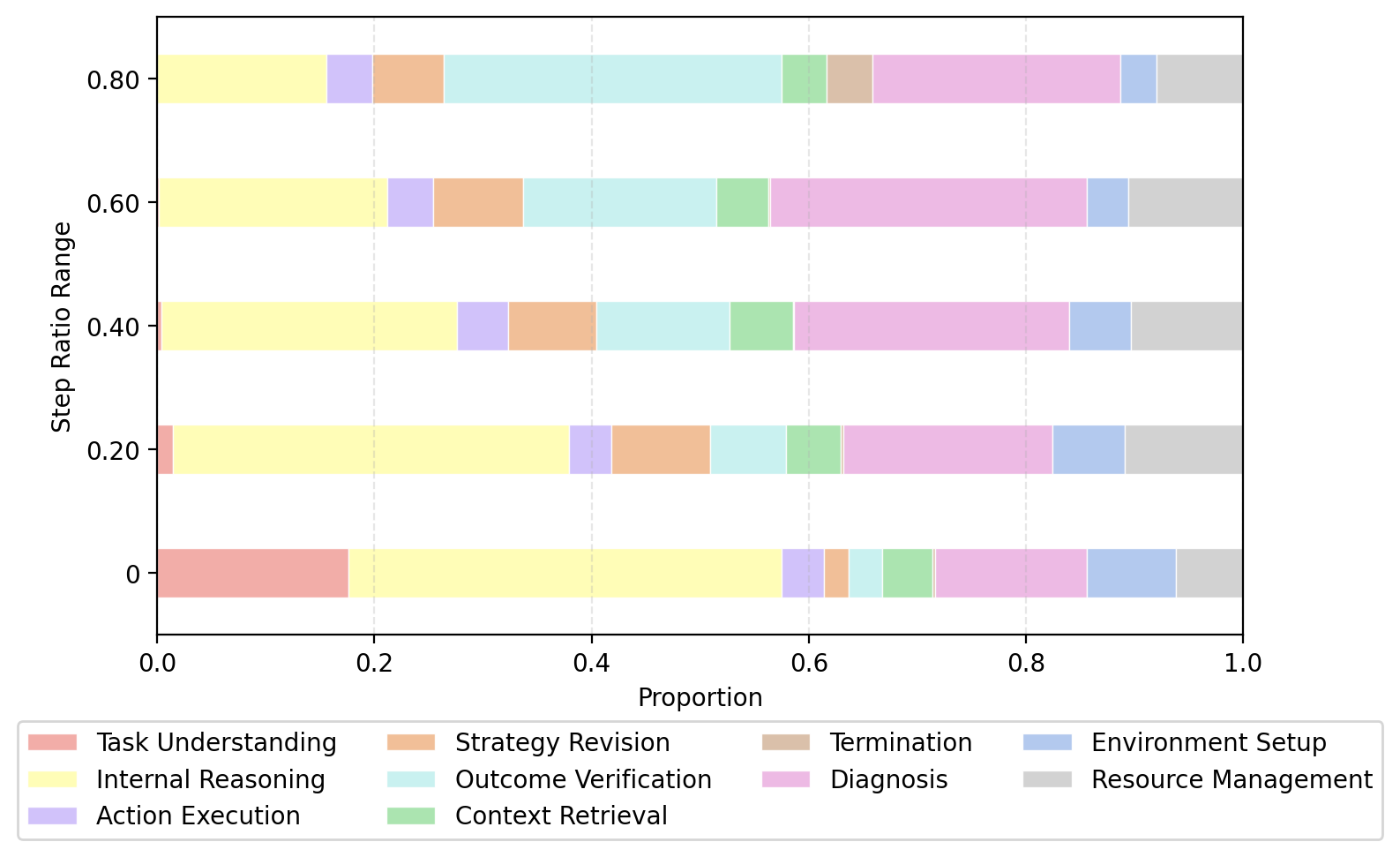}
    \caption{DeepSeek-V3.2 -- Terminus~2}
  \end{subfigure}

  \vspace{1mm}
  \begin{subfigure}[t]{\figGridRangeW}
    \centering
    \includegraphics[width=\linewidth]{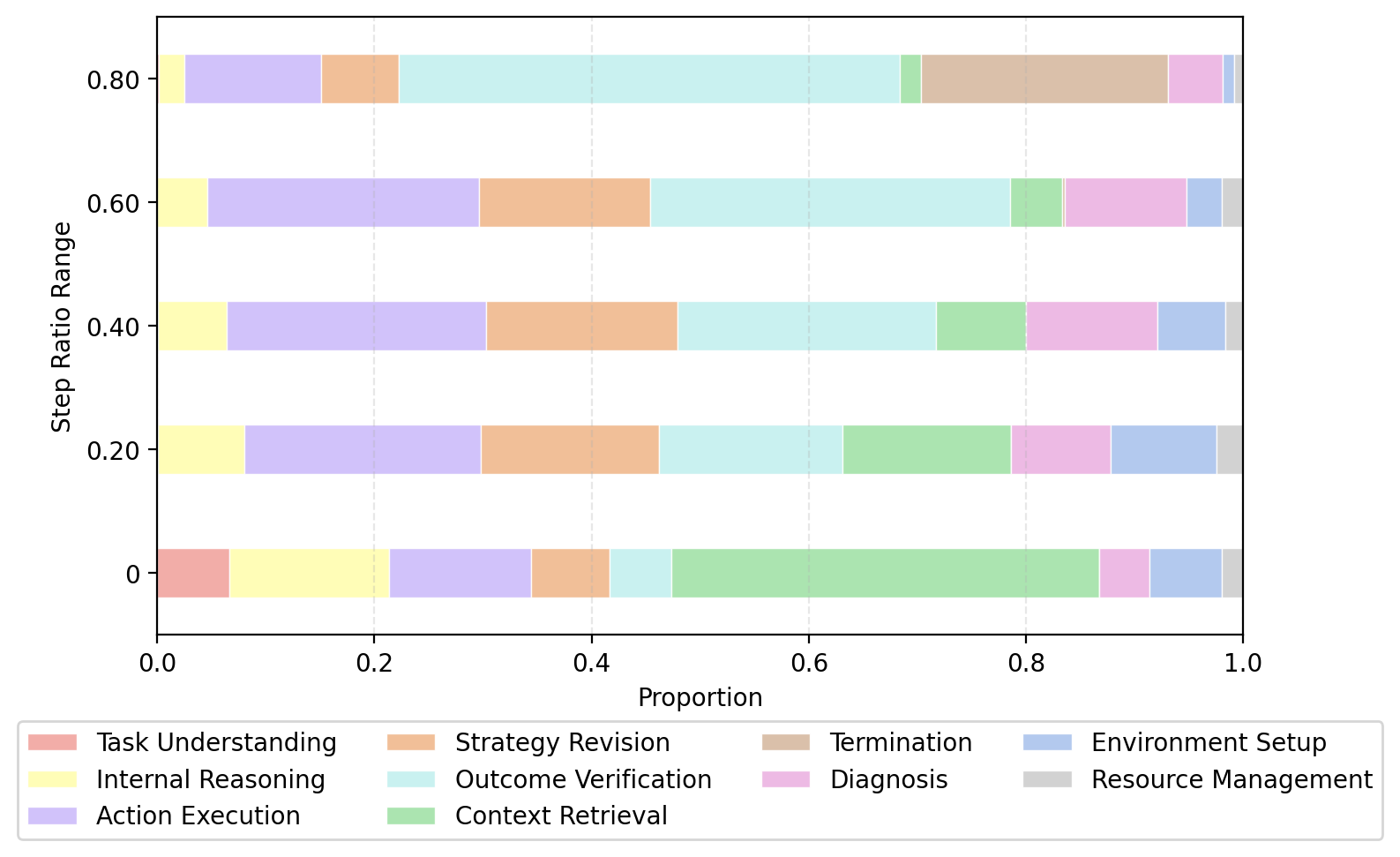}
    \caption{Qwen3-Coder-480B -- MiniSWE}
  \end{subfigure}\hfill
  \begin{subfigure}[t]{\figGridRangeW}
    \centering
    \includegraphics[width=\linewidth]{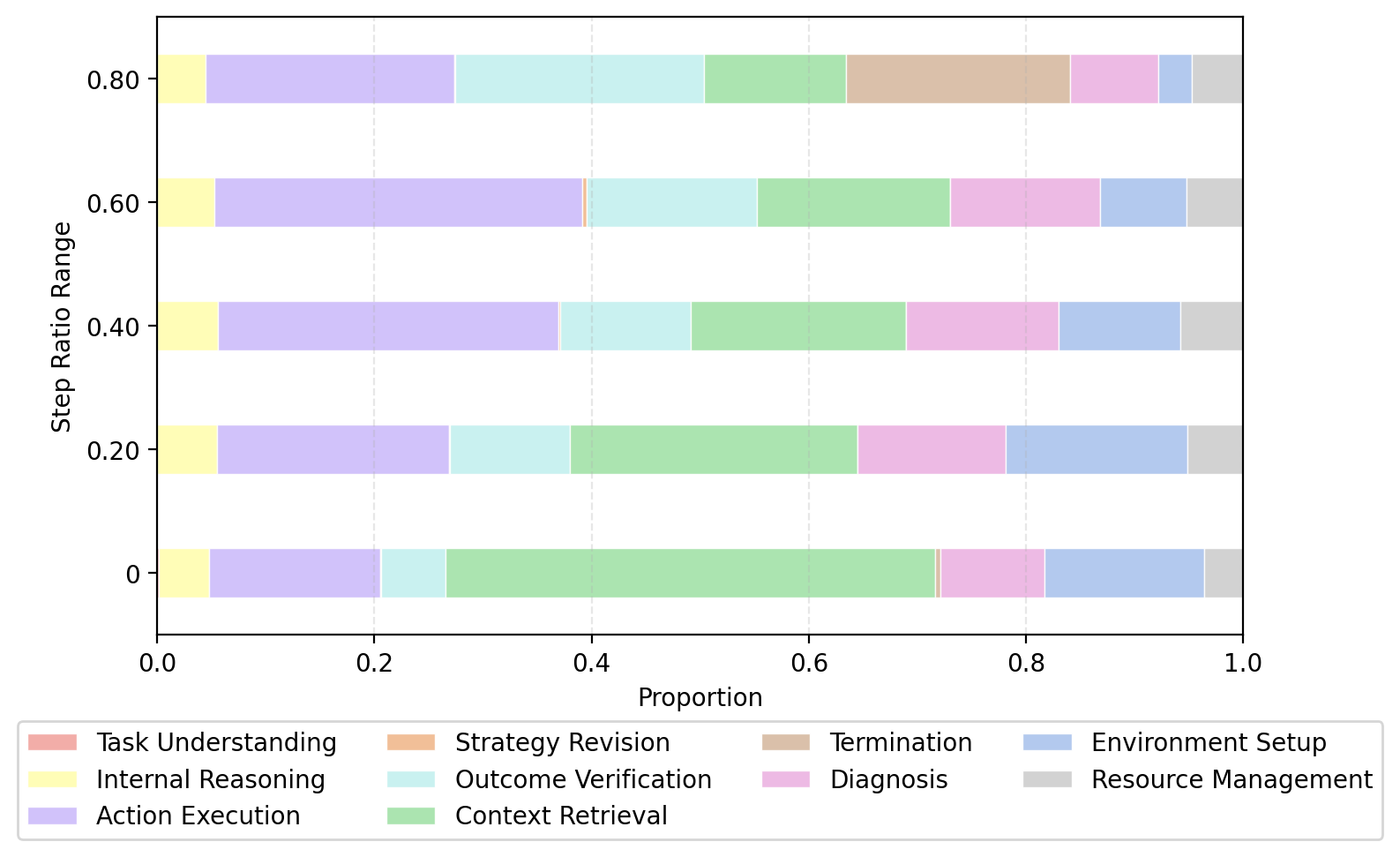}
    \caption{Qwen3-Coder-480B -- OpenHands}
  \end{subfigure}\hfill
  \begin{subfigure}[t]{\figGridRangeW}
    \centering
    \includegraphics[width=\linewidth]{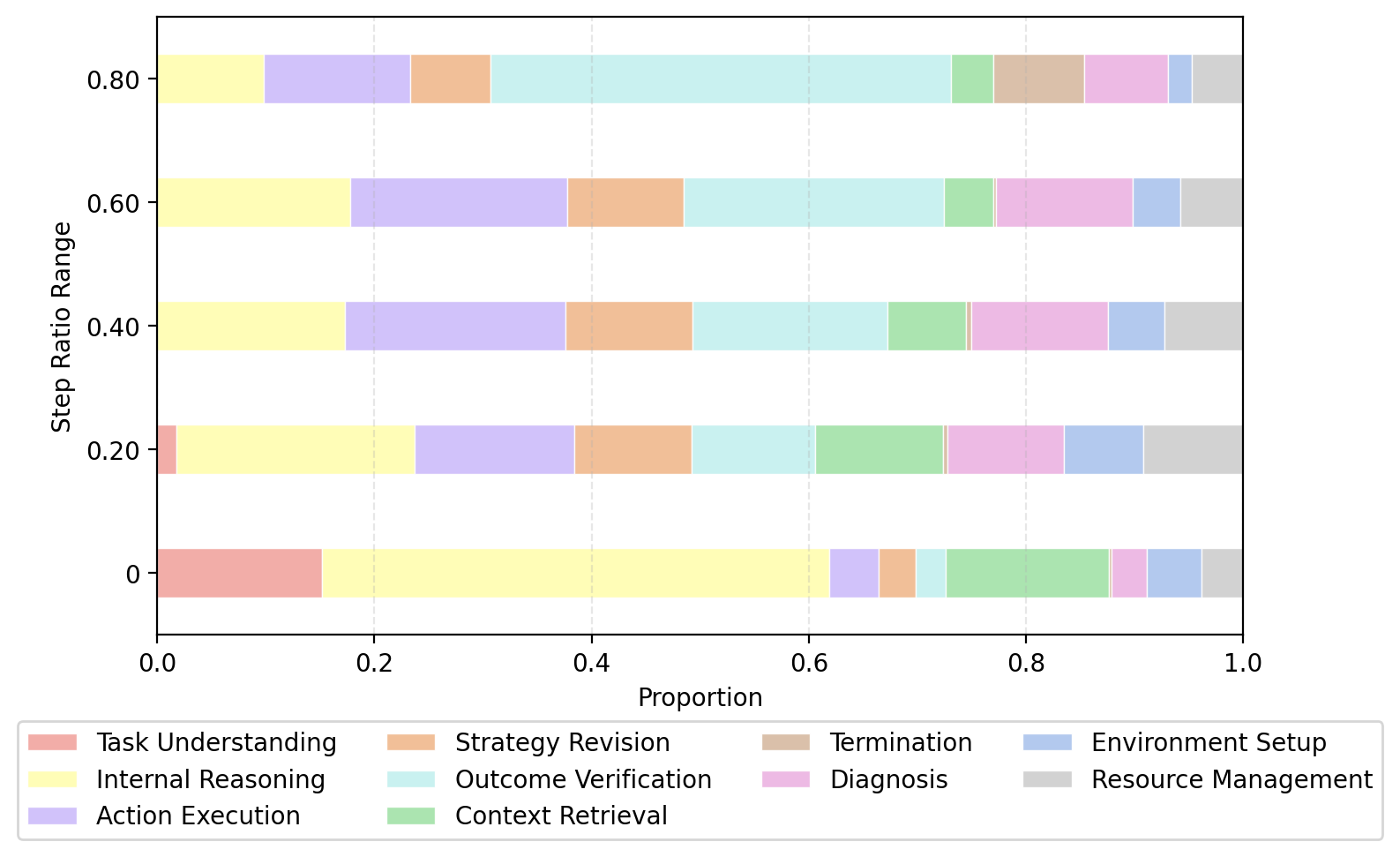}
    \caption{Qwen3-Coder-480B -- Terminus~2}
  \end{subfigure}

  \vspace{1mm}
  \begin{subfigure}[t]{\figGridRangeW}
    \centering
    \includegraphics[width=\linewidth]{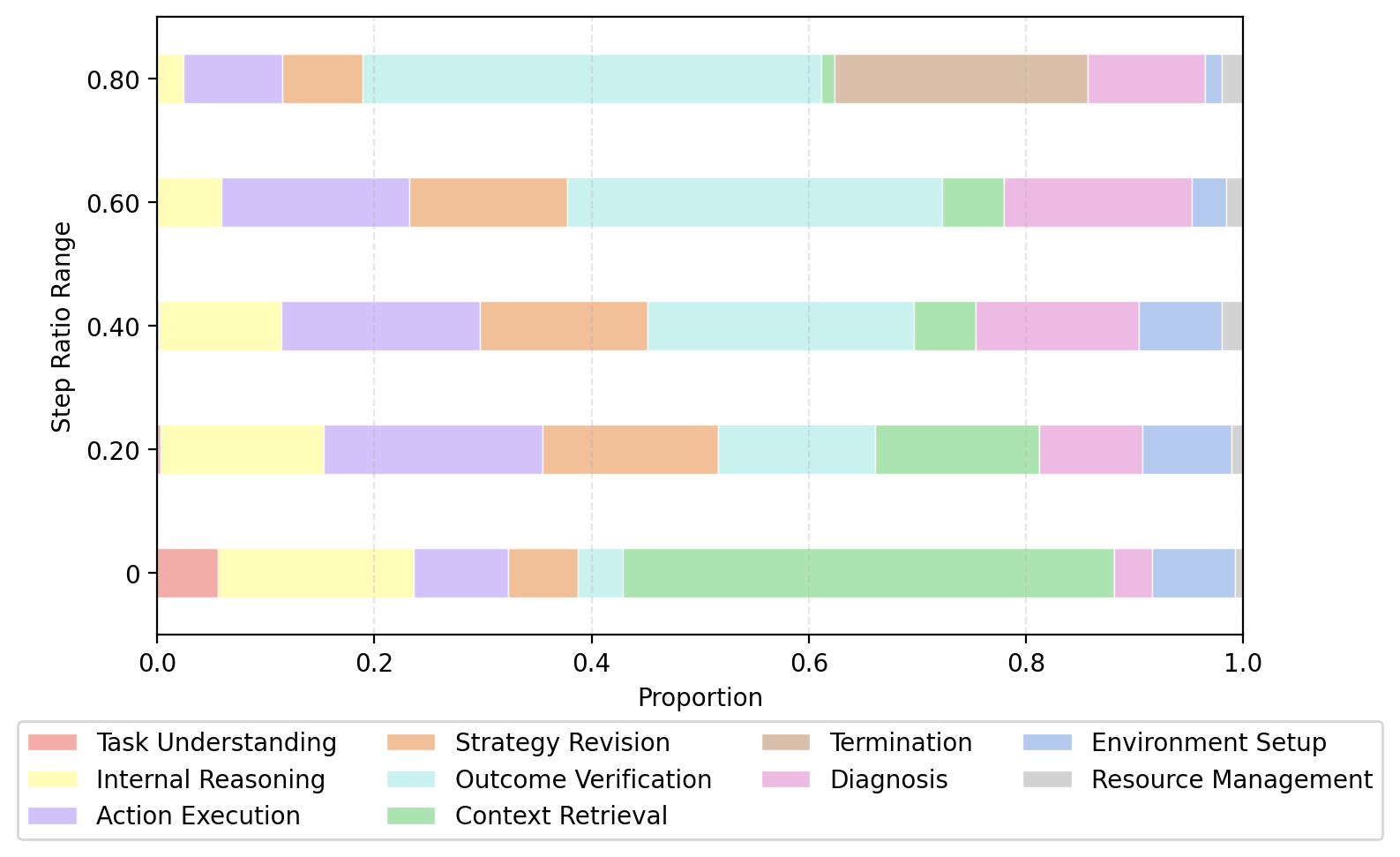}
    \caption{Kimi-K2-Instruct -- MiniSWE}
  \end{subfigure}\hfill
  \begin{subfigure}[t]{\figGridRangeW}
    \centering
    \includegraphics[width=\linewidth]{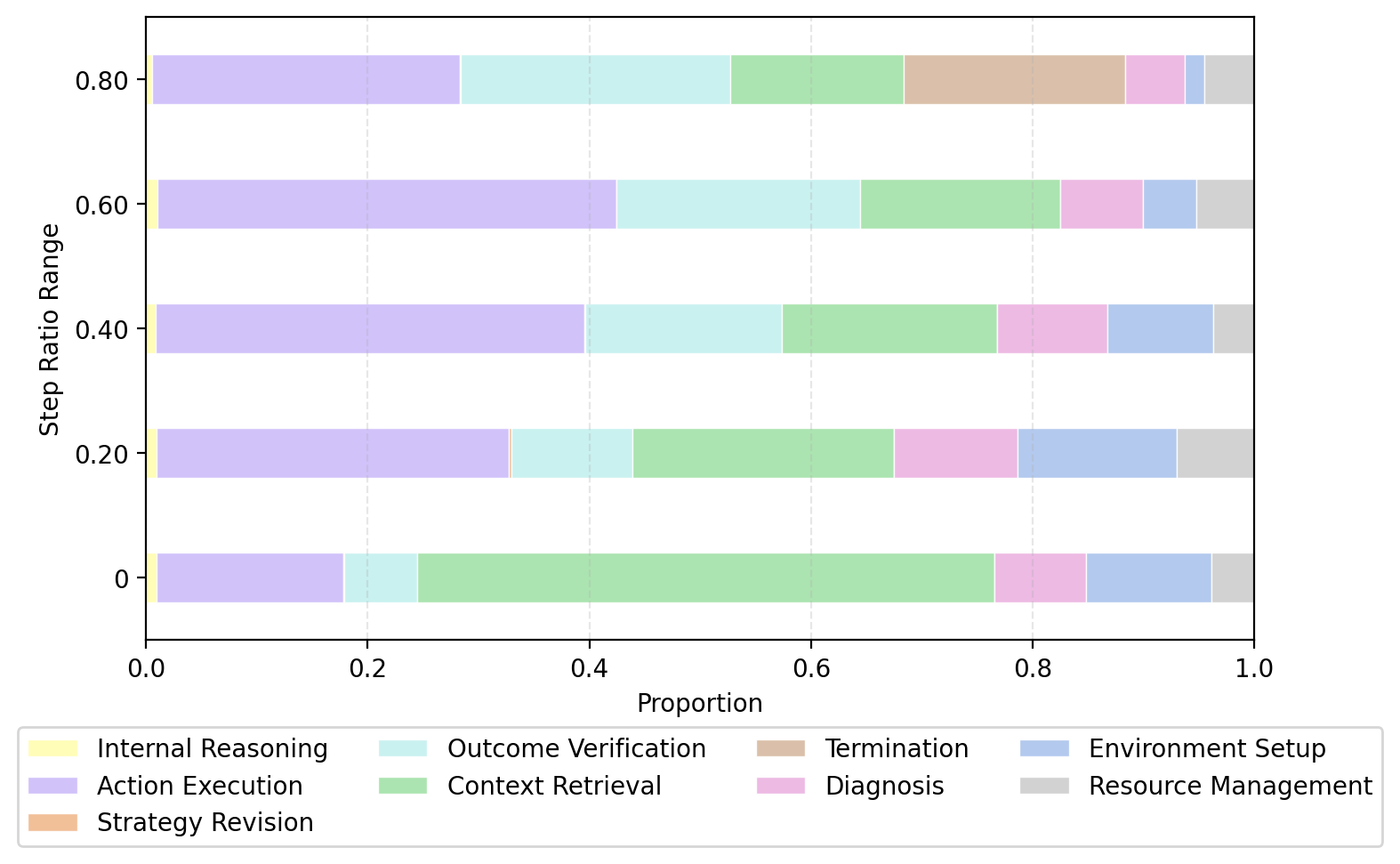}
    \caption{Kimi-K2-Instruct -- OpenHands}
  \end{subfigure}\hfill
  \begin{subfigure}[t]{\figGridRangeW}
    \centering
    \includegraphics[width=\linewidth]{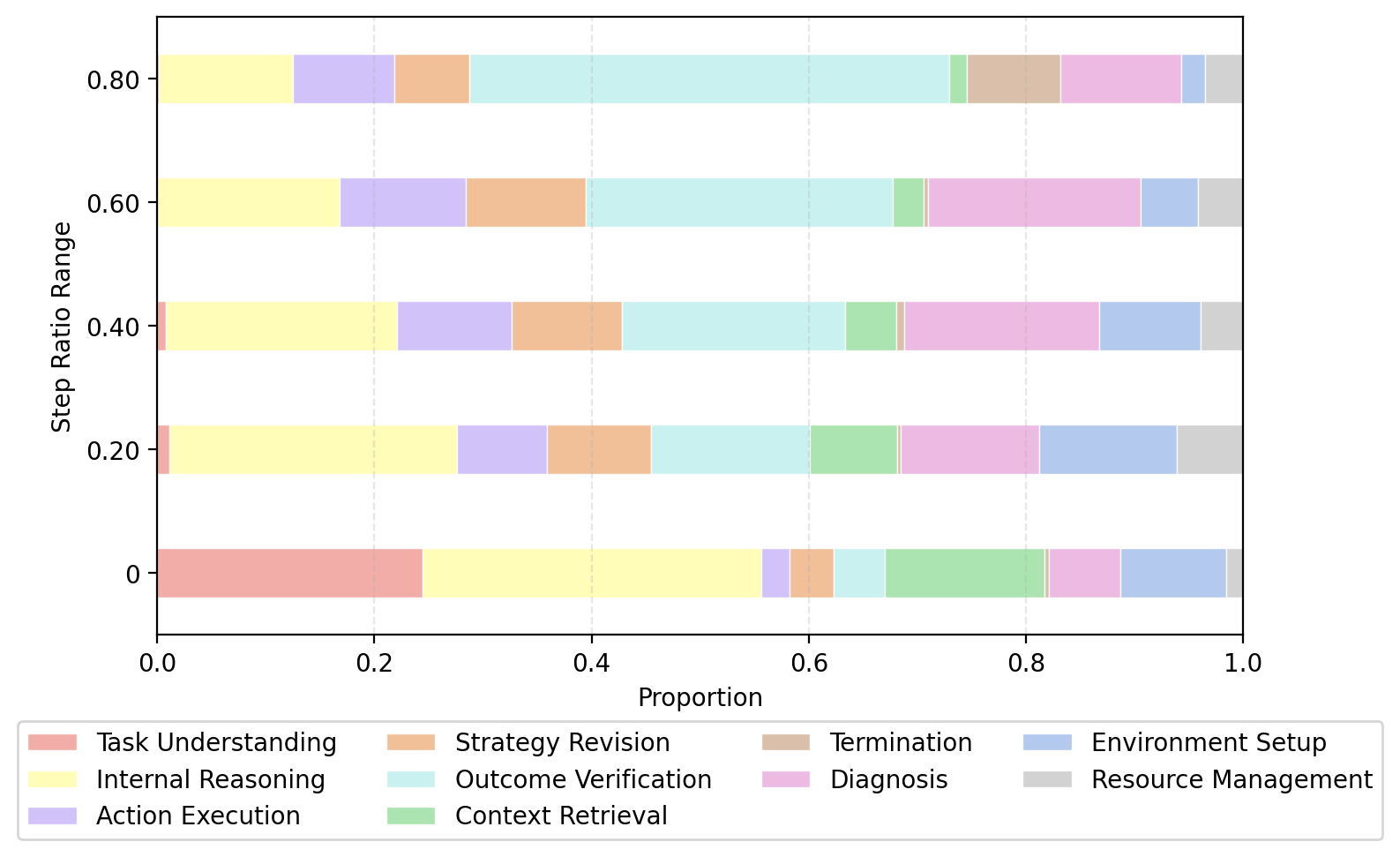}
    \caption{Kimi-K2-Instruct -- Terminus~2}
  \end{subfigure}

  \caption{\textbf{Step-range error distribution for all 15 backbone--agent combinations.}
  Rows correspond to backbones (Claude-sonnet-4, GPT-5, DeepSeek-V3.2, Qwen3-Coder-480B, Kimi-K2-Instruct);
  columns correspond to agent frameworks (MiniSWE-Agent, OpenHands, Terminus~2).
  Each panel shows the stacked area ratio of error critical steps across execution stages, 
  illustrating where in the workflow failures tend to concentrate for each pairing.}
  \label{fig:step_range_grid}
\end{figure}

\parabf{Step range error distribution across model backbones.}
\Cref{fig:step_range_grid} presents the complete $5 \times 3$ grid of stacked area ratio distributions across execution stages.
Each row fixes a backbone and each column fixes an agent framework, covering all 15 combinations.
The visualizations reveal how error critical steps concentrate at different phases of the agent workflow, with notable variations across backbone--agent pairings.

\section{Industrial Agent Analysis: Claude Code}
\label{sec:app:claude_code}

We apply \system to analyze trajectories generated by Claude Code, an industrial coding agent, and compare its architecture and behavior with the academic agent frameworks studied in the main text.

\parabf{Architecture.}
Claude Code's source comprises 40+ specialized tools organized into eight categories (file operations, shell execution, search and navigation, agent orchestration and planning, web and external services, workspace configuration, task management, and others), compared to 5--10 tools in typical academic agents.
The system includes dedicated modules for context compaction, token budgeting, and feature-gated code paths that are absent from research frameworks.

\parabf{Key findings.}
Our analysis reveals several structural differences between industrial and academic agents:
\begin{enumerate}
  \item \textbf{Tooling investment.} Industrial agents invest heavily in specialized tooling and error recovery infrastructure, while academic agents operate with a narrow, general-purpose tool set.
  \item \textbf{Context management.} Production agents implement sophisticated context management (compaction, budget tracking, feature gating) that academic agents typically lack, enabling longer effective trajectories.
  \item \textbf{Exploration-to-change ratio.} The exploration-to-change ratio is a strong predictor of trajectory quality: Claude Code exhibits a lower ratio (more actions per exploration step) that correlates with higher task success.
  \item \textbf{Parallel execution.} Parallel tool execution, available in industrial agents, significantly reduces wall-clock time but introduces ordering-sensitivity issues absent from sequential academic frameworks.
  \item \textbf{RL feedback signals.} Per-step deviation labels produced by \system on industrial agent trajectories can serve as dense training signals, potentially bridging the behavioral gap between industrial and academic agents.
\end{enumerate}

Note that because the Claude Code trajectories were collected on a different task distribution (TerminalBench tasks executed via the Claude Code CLI), direct numerical comparison with the academic agent results in the main text is not straightforward. The findings above therefore focus on qualitative architectural and behavioral differences rather than aggregate metric comparisons.

\end{document}